\documentclass[letterpaper,aps,superscriptaddres]{revtex4}
\usepackage{graphicx, color, amsmath,amssymb,latexsym,cleveref}

\def\lsim{\mathrel{\raise.3ex\hbox{$<$\kern-.75em\lower1ex\hbox{$\sim$}}}}
\def\gsim{\mathrel{\raise.3ex\hbox{$>$\kern-.75em\lower1ex\hbox{$\sim$}}}}

\newcommand{\mini}[2]{\begin{minipage}{#1}\centering #2\end{minipage}}

\begin{document}

\title{Indications of Negative Evolution for the Sources of the Highest Energy Cosmic Rays}



\author{Andrew~M.~Taylor}
\affiliation{Dublin Institute for Advanced Studies, 31 Fitzwilliam Place, Dublin 2, Ireland\vspace{-\parskip}}

\author{Markus~Ahlers} 
\affiliation{Wisconsin IceCube Particle Astrophysics Center (WIPAC), Madison, WI 53703, USA\vspace{-\parskip}}
\affiliation{Department of Physics, University of Wisconsin-Madison, Madison, WI 53706, USA}

\author{Dan~Hooper} 
\affiliation{Center for Particle Astrophysics, Fermi National Accelerator Laboratory, Batavia, IL 60510, USA\vspace{-\parskip}}
\affiliation{Department of Astronomy and Astrophysics, University of Chicago, Chicago, IL 60637}


\begin{abstract}

Using recent measurements of the spectrum and chemical composition of the highest energy cosmic rays, we consider the sources of these particles. We find that the data strongly prefers models in which the sources of the ultra-high energy cosmic rays inject predominantly intermediate mass nuclei, with comparatively few protons or heavy nuclei, such as iron or silicon. If the number density of sources  per comoving volume does not evolve with redshift, the injected spectrum must be very hard ($\alpha \simeq 1$) in order to fit the spectrum observed at Earth. Such a hard spectral index would be surprising and difficult to accommodate theoretically. In contrast, much softer spectral indices, consistent with the predictions of Fermi acceleration ($\alpha \simeq 2$), are favored in models with negative source evolution. With this theoretical bias, these observations thus favor models in which the sources of the highest energy cosmic rays are preferentially located within the low-redshift universe.
\end{abstract}

\hspace{5.2in} \mbox{FERMILAB-PUB-15-235-A}

\maketitle

\section{Introduction}
\label{intro}

Despite considerable advances in the measurement of the spectra, arrival directions, and chemical composition of the ultra-high energy cosmic rays (UHECRs), we remain ignorant of the sources of these particles. In particular, although it had been hoped that anisotropies in the arrival directions of the UHECRs would be detectable by the Pierre Auger Observatory (PAO), such studies have thus far not conclusively identified a correlation between these particles and any known class of astrophysical objects~\cite{PierreAuger:2014yba}. Efforts to identify UHECR sources using secondary signatures have also been inconclusive~\cite{Aab:2014bha,Abreu:2013zbq}. No signal capable of revealing the origin of the UHECRs has yet appeared. 

Despite the lack of such a definitive result, the available information bearing on this puzzle has significantly increased over the past few years, allowing for considerable progress to be made. Of particular importance are the PAO's measurements of the chemical composition of the UHECR spectrum~\cite{Aab:2014kda,Aab:2014aea}, which reveal the highest energy cosmic rays to be dominated by intermediate mass nuclei, as opposed to either protons or heavy nuclei, such as iron or silicon. While earlier data also supported similar conclusions~\cite{Hooper:2009fd}, the PAO's most recent template-based composition study~\cite{Aab:2014aea} has made a rather detailed determination possible.

Considering the most recent spectrum and composition measurements from the PAO~\cite{Aab:2014kda} above an energy threshold of $10^{18.6}$~eV, we investigate models for the injected spectrum, composition, and redshift distribution of the UHECR sources. Our main finding is that in models with no significant evolution with redshift (a constant number of sources per comoving volume), the injected spectrum from the sources of the UHECRs must exhibit a very hard spectral index, $\alpha \simeq 1.1$. Such a hard spectrum would be surprising and difficult to accommodate theoretically (although scenarios have been proposed~\cite{1988MNRAS.235..997H,Blasi:2000xm,Fang:2013cba,Unger:2015laa,Globus:2015xga}). Models with positive redshift evolution (a greater number of sources per comoving volume at high redshift) only exacerbate this problem. 
In models with significant negative evolution, however, the measured spectrum can be accommodated for more well-motivated values of the injected spectral index, near that generically predicted from Fermi acceleration, $\alpha \simeq 2$.  This theoretical bias for soft cosmic ray injection spectra supports the conclusion that the sources of the UHECRs are predominantly located in the low-redshift universe. One notable implication of this result is that a low cosmogenic neutrino flux is predicted, well below the expected sensitivity of existing or planned experiments. 

In this paper, we investigate the propagation of ultra-high energy cosmic ray nuclei and compare the predictions of various models to the current data from the PAO.
In Sec.~\ref{propagation}, we describe our treatment of UHECR nuclei propagation and the range of models considered. In Sec.~\ref{results}, we describe our results, which favor models with negative redshift evolution and which inject significant quantities of intermediate mass nuclei, such as helium, carbon, nitrogen, and oxygen. In Sec.~\ref{constraints}, we discuss the contribution from UHECR propagation to the diffuse gamma-ray background and the flux of cosmogenic neutrinos. Lastly, in Sec.~\ref{conclusion}, we summarize our results and conclude.

\section{The Propagation of Ultra-High Energy Cosmic Ray Nuclei}
\label{propagation}

To test various models for the sources of the highest energy cosmic rays, we employ a Monte Carlo description of UHECR propagation, as originally described in Ref.~\cite{Hooper:2006tn}. In this calculation, UHECR protons and nuclei 
are propagated through the cosmic microwave background (CMB) and the cosmic infrared background (CIB), undergoing photo-disintegration, photo-pion interactions, pair production, and redshift energy losses. In order to allow for a large number of models to be studied quickly, 
we inject particles with energies and distances over discretized logarithmic ranges, and then sum these results with appropriate weighting factors, so as to generate results for more general functions of the source energy spectra and spatial distribution.

For each model, we calculate the average depth of shower maximum, $\langle X_{\rm max} \rangle$, and its RMS variation, RMS$(X_{\rm max})$, as a function of energy, and compare this to the values measured by the PAO~\cite{Aab:2014kda}.  In order to encapsulate the uncertainty associated with the modeling of the shower development, we plot a band which encompasses the values as determined using the different hadronic models, EPOS-LHC~\cite{Pierog:2013ria} and QGSJET-II-04~\cite{Ostapchenko:2010vb}, which have each been updated to incorporate the first results from the Large Hadron Collider, as well as SIBYLL~\cite{Ahn:2009wx}. The cross sections and target photon spectral distributions relevant 
for proton energy losses are sufficiently well understood so as not to introduce
large uncertainties in our results. Furthermore, although the uncertainties regarding our knowledge of 
the photo-disintegration cross sections and the CIB spectral distribution relevant for
nuclei propagation are not necessarily negligible, such uncertainties are unlikely to qualitatively impact our conclusions. In our calculations, we adopt the cross sections as given in Ref.~\cite{TALYS} and the CIB spectral distribution as described in Ref.~\cite{Franceschini:2008tp}.

In this study, we consider models with a wide range of redshift distributions, energy spectra, and chemical compositions. Given our present ignorance of the nature of these sources, we consider a simple source evolution model with the aim to encapsulate a broad set of possible histories, which we parameterize by: 
\begin{eqnarray}
\frac{{\rm d}N}{{\rm d}V_{C}} \propto (1+z)^{n}, \hspace{1cm}z<z_{\rm max},
\label{evolutioneq}
\end{eqnarray}
where ${\rm d}N/{\rm d}V_C$ is the number of sources per comoving volume. We will present results for a wide range of evolution models, from $n=3$ to $n=-6$. We adopt a maximum redshift value of $z_{\rm max}=3$, which is well beyond the distance that UHECRs are able to propagate. 

For the spectra of UHECRs injected from sources, we adopt the following form:
\begin{eqnarray}
\frac{{\rm d}N}{{\rm d}E}\propto E^{-\alpha}\exp[-E/E_{\rm Z, max}],
\label{source_spec}
\end{eqnarray}
where $\alpha$ describes the spectral index and $E_{\rm Z, max}=(Z/26) \times E_{\rm Fe,max}$ is a rigidity dependent maximum energy. 

For the chemical composition of the UHECR spectrum at injection, we consider an arbitrary mixture of 5 nuclear species, namely:  protons, helium, nitrogen, silicon, and iron. Physically, we take these categories to represent groups of nuclei. For example, the nitrogen fraction can be thought of as a proxy for the fraction of injected cosmic rays that are carbon, nitrogen, or oxygen.

\begin{figure}[h!]
\centering\leavevmode
\includegraphics[height=0.25\linewidth, viewport =  40 46 750 556,clip=true]{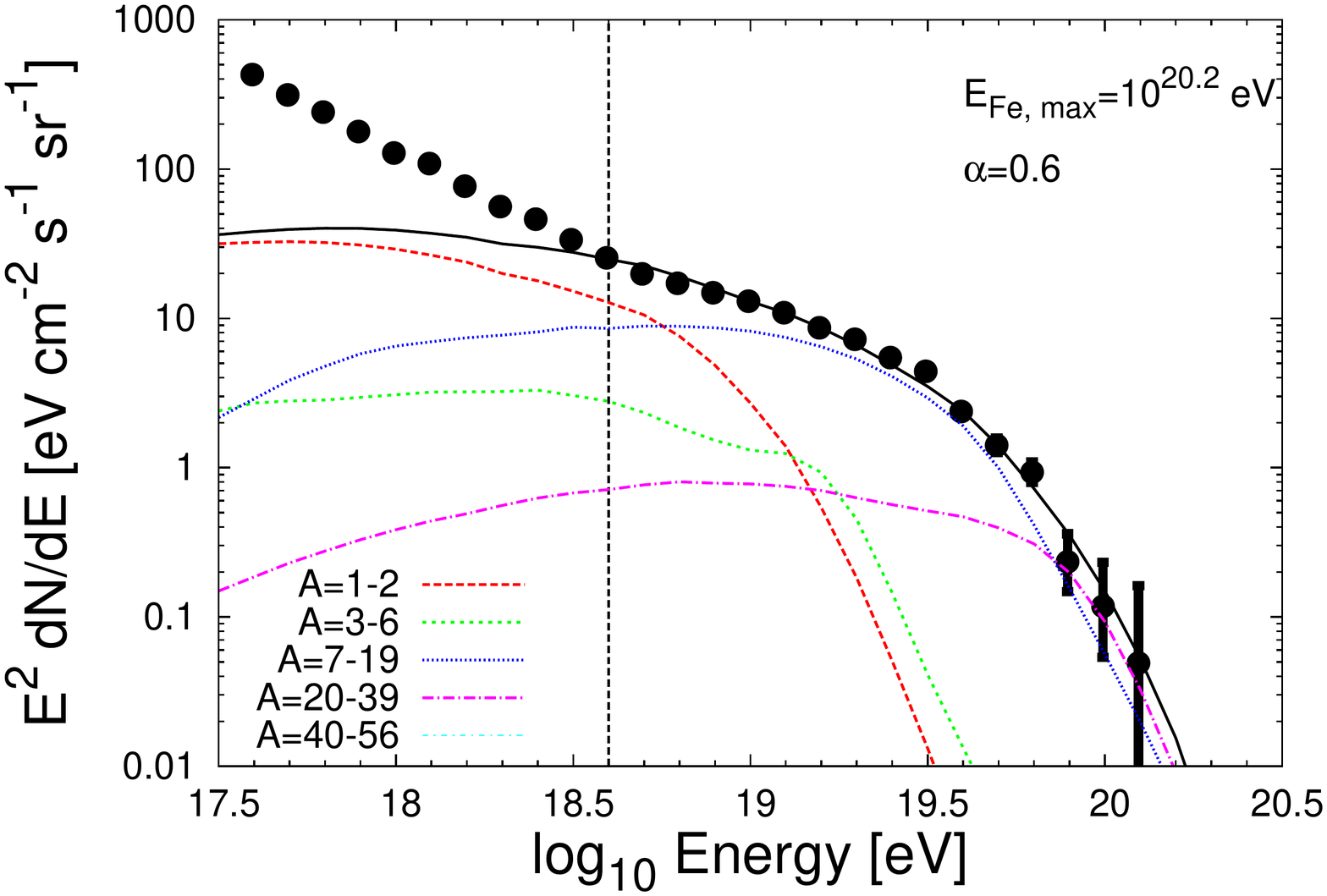}\includegraphics[height=0.25\linewidth, viewport =  0 -32 650 468,clip=true]{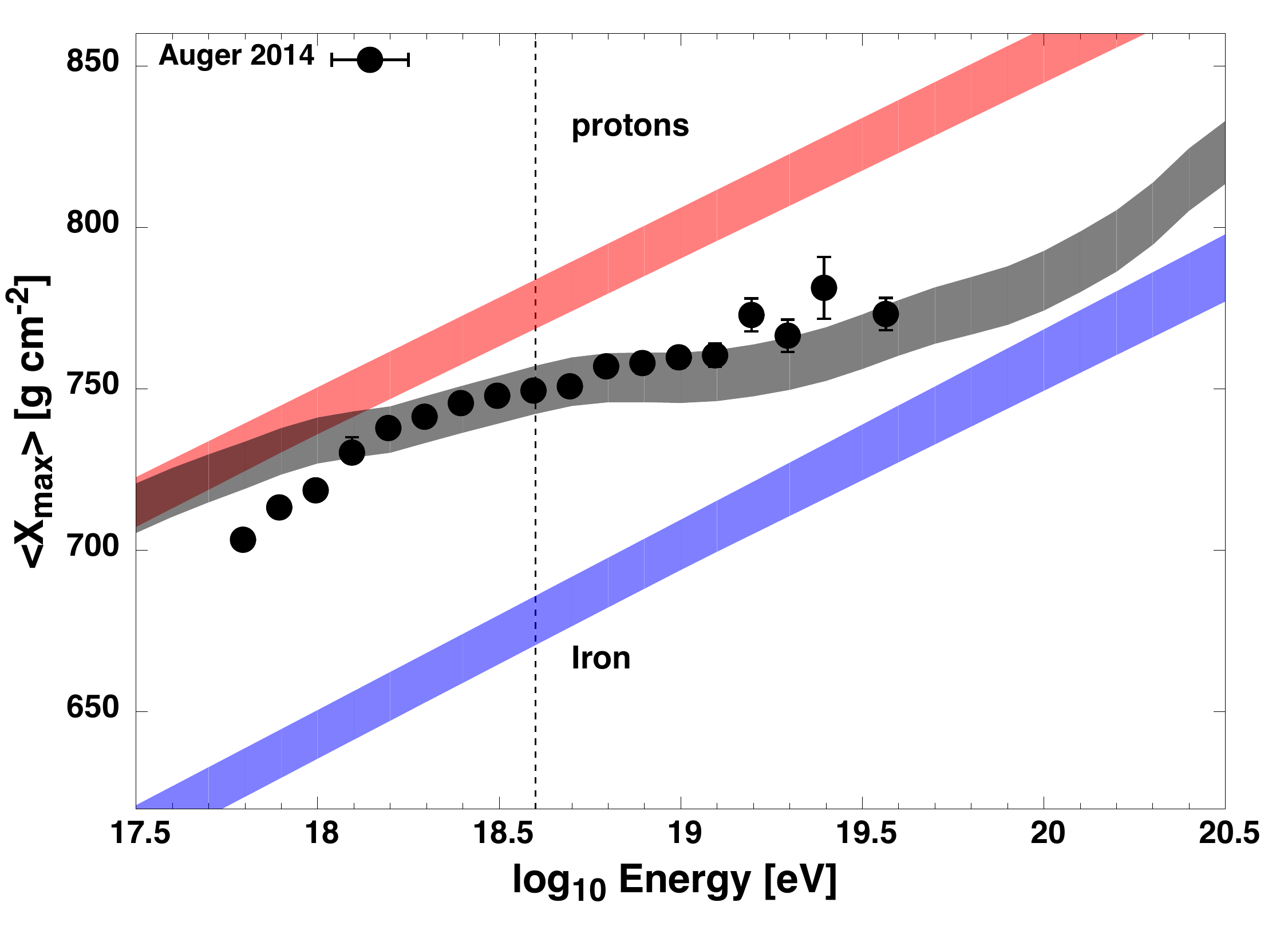}\includegraphics[height=0.25\linewidth, viewport =  0 -32 650 468,clip=true]{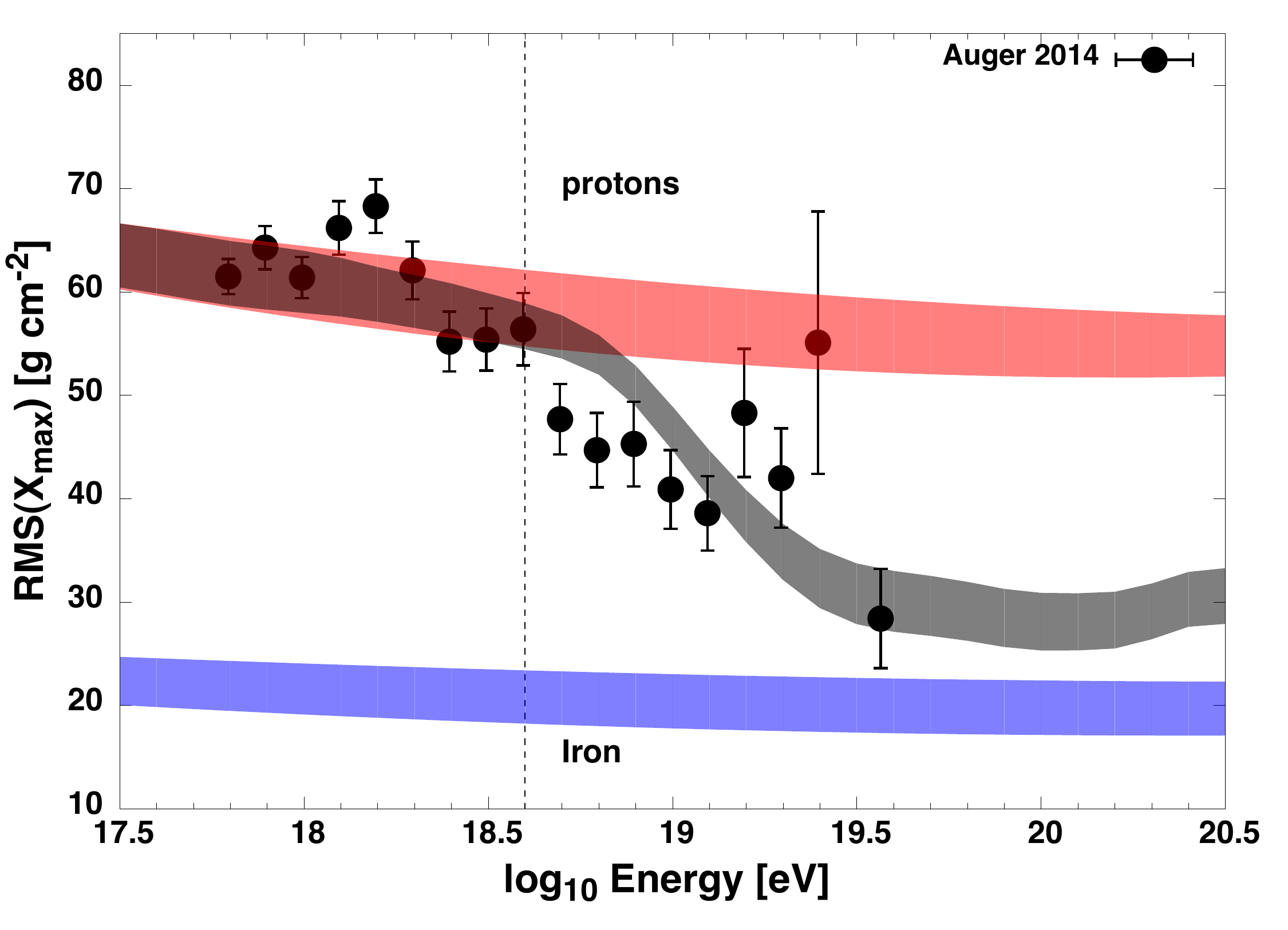}\\
\includegraphics[height=0.25\linewidth, viewport =  40 46 750 556,clip=true]{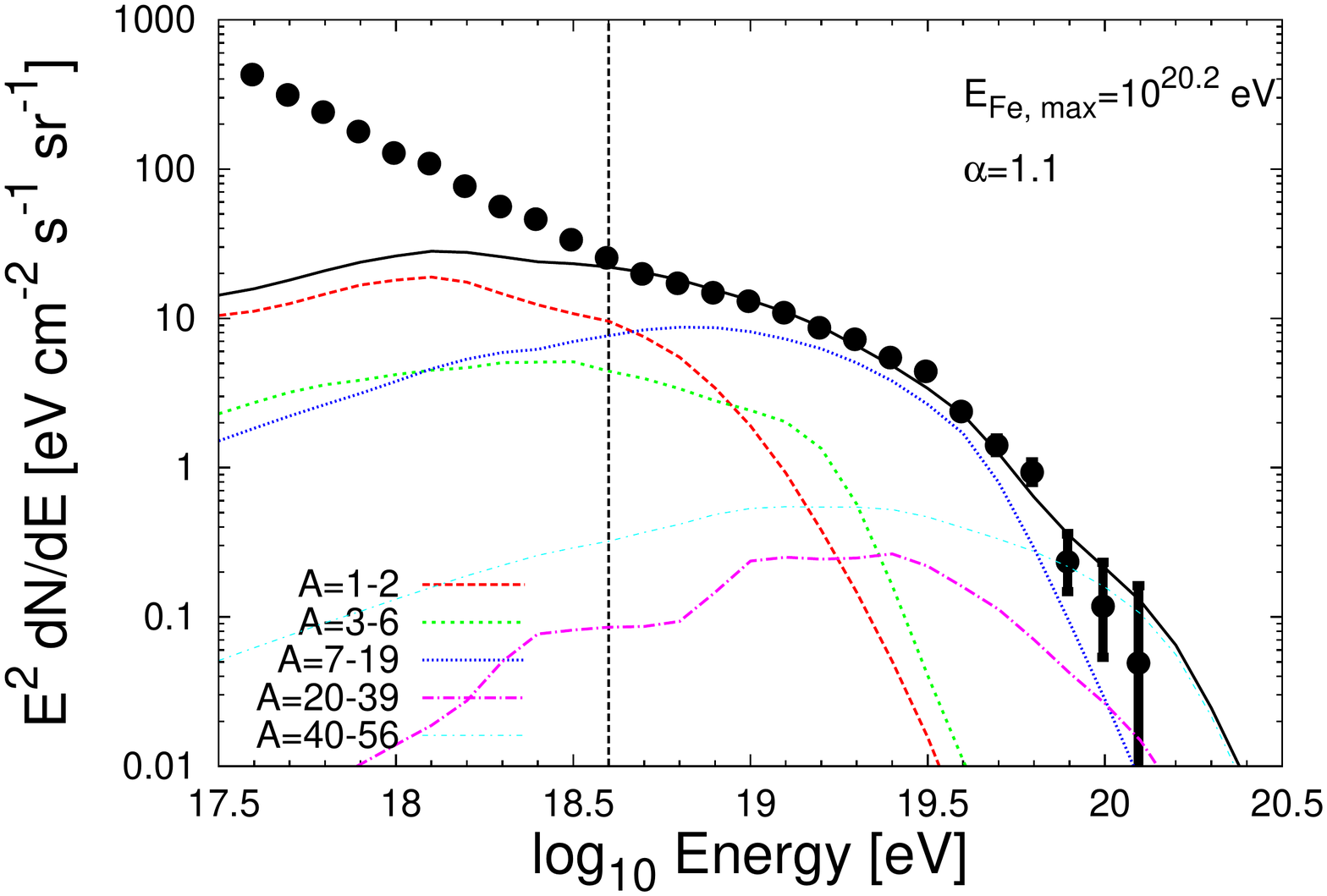}\includegraphics[height=0.25\linewidth, viewport =  0 -32 650 468,clip=true]{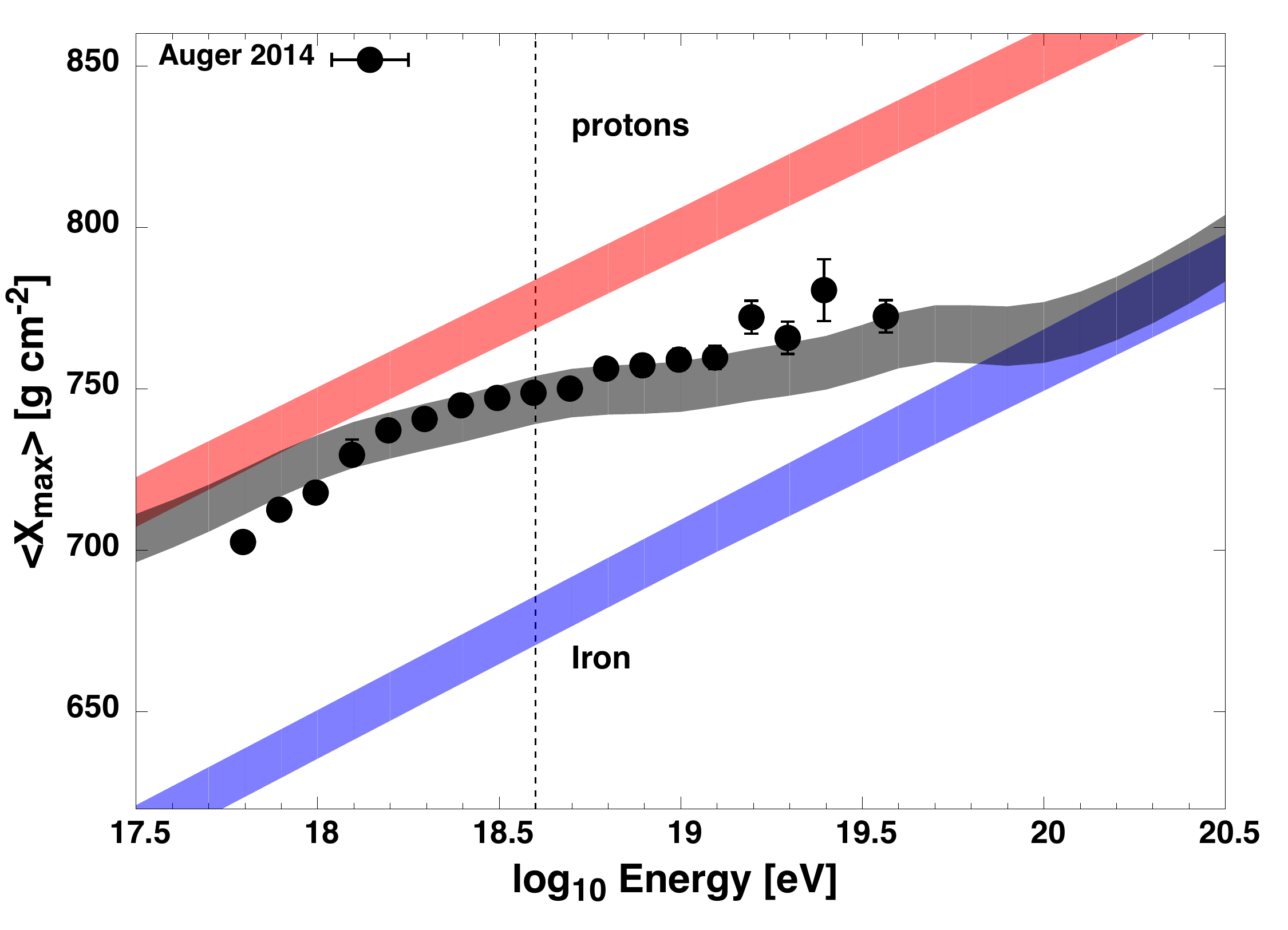}\includegraphics[height=0.25\linewidth, viewport =  0 -32 650 468,clip=true]{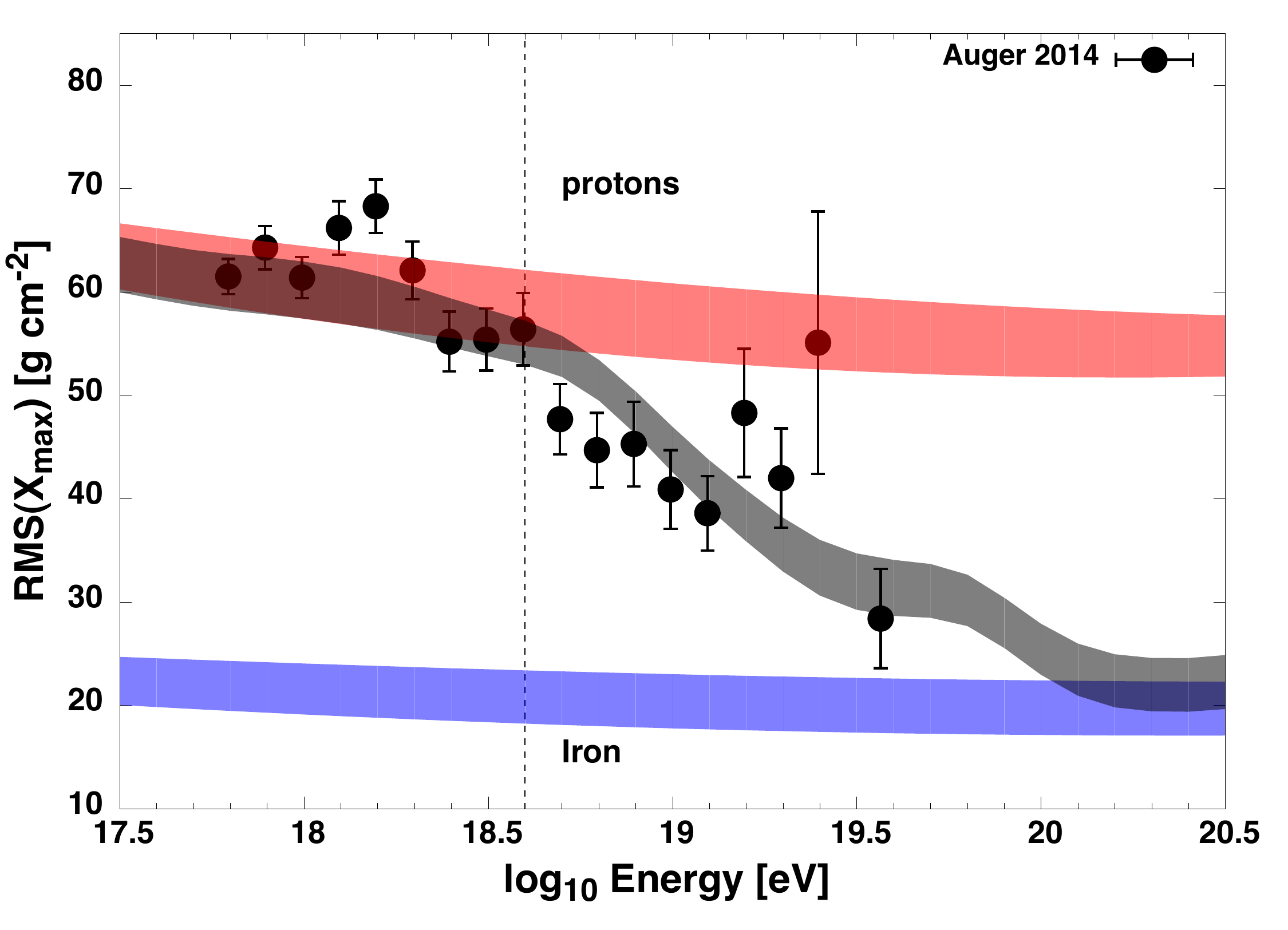} \\
\includegraphics[height=0.25\linewidth, viewport =  40 46 750 556,clip=true]{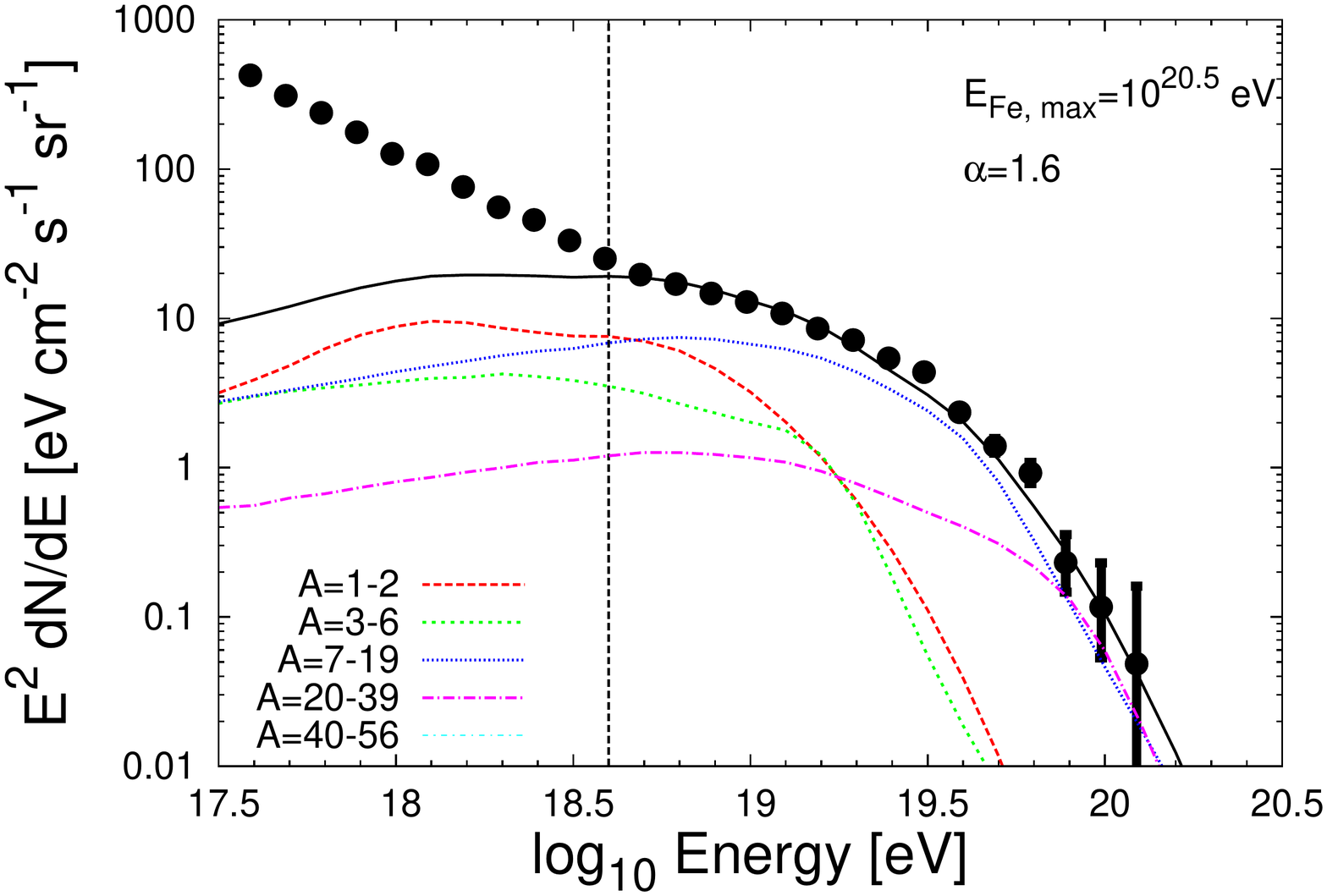}\includegraphics[height=0.25\linewidth, viewport =  0 -32 650 468,clip=true]{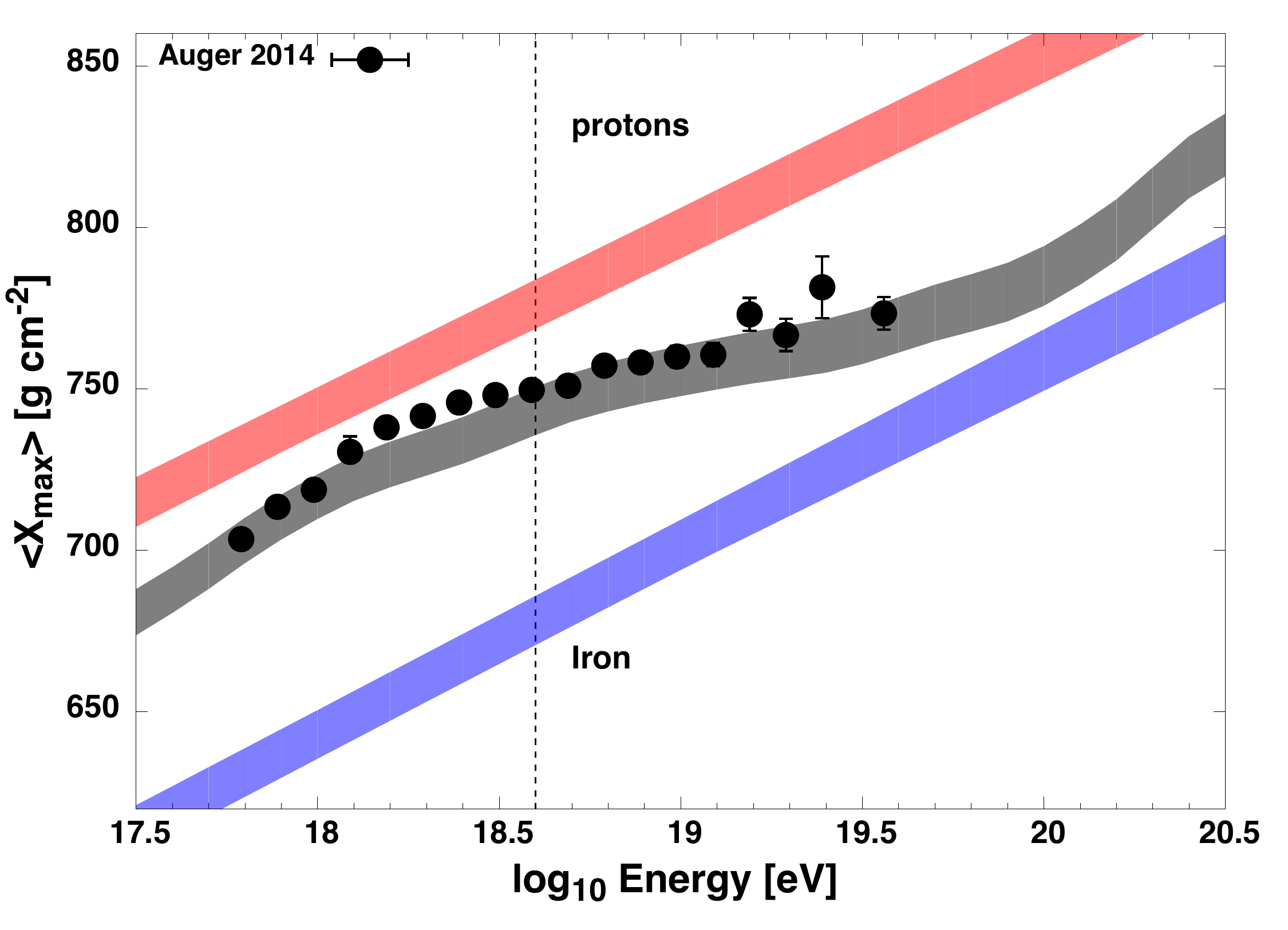}\includegraphics[height=0.25\linewidth, viewport =  0 -32 650 468,clip=true]{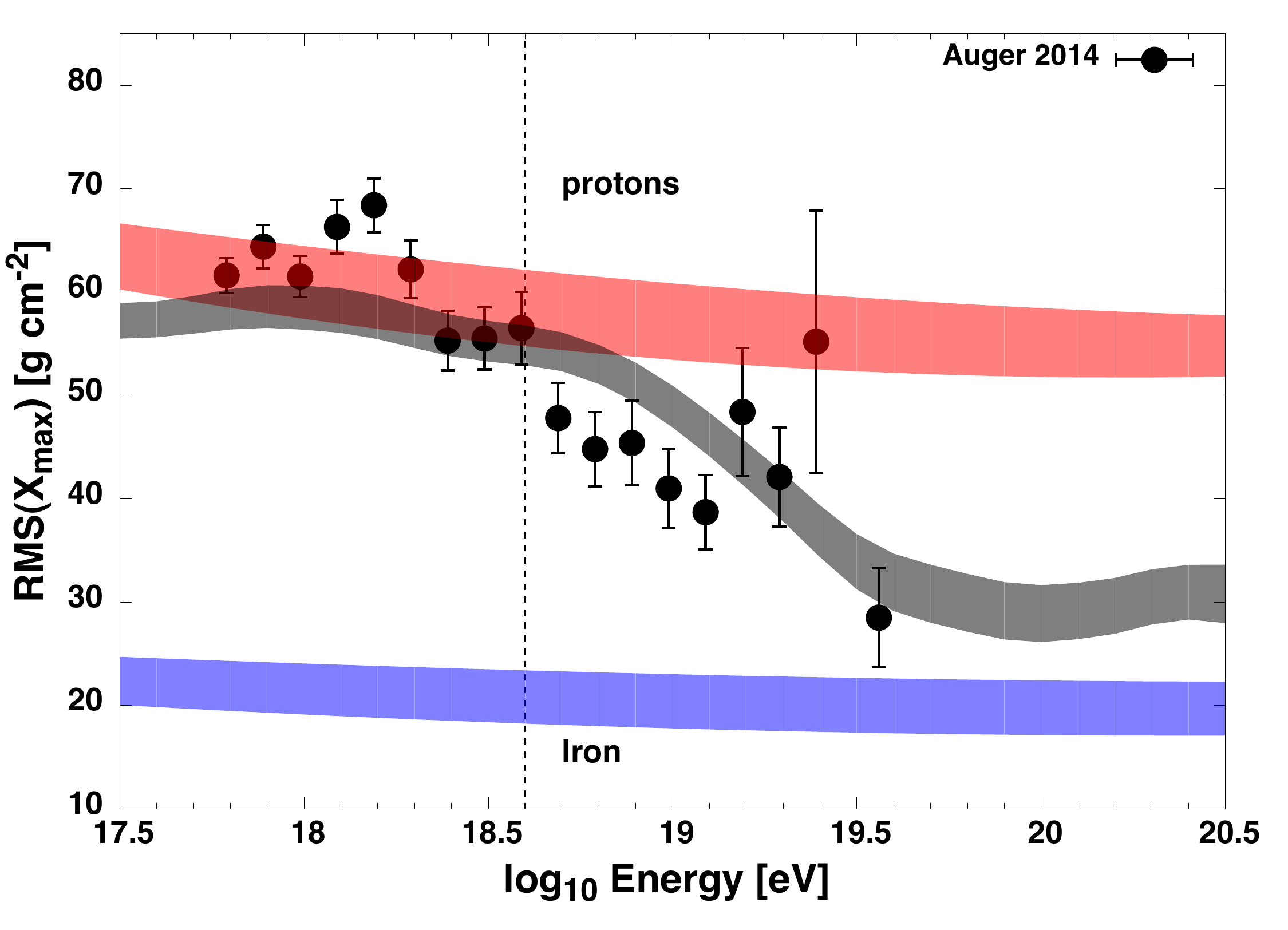}\\
\includegraphics[height=0.25\linewidth, viewport =  40 46 750 556,clip=true]{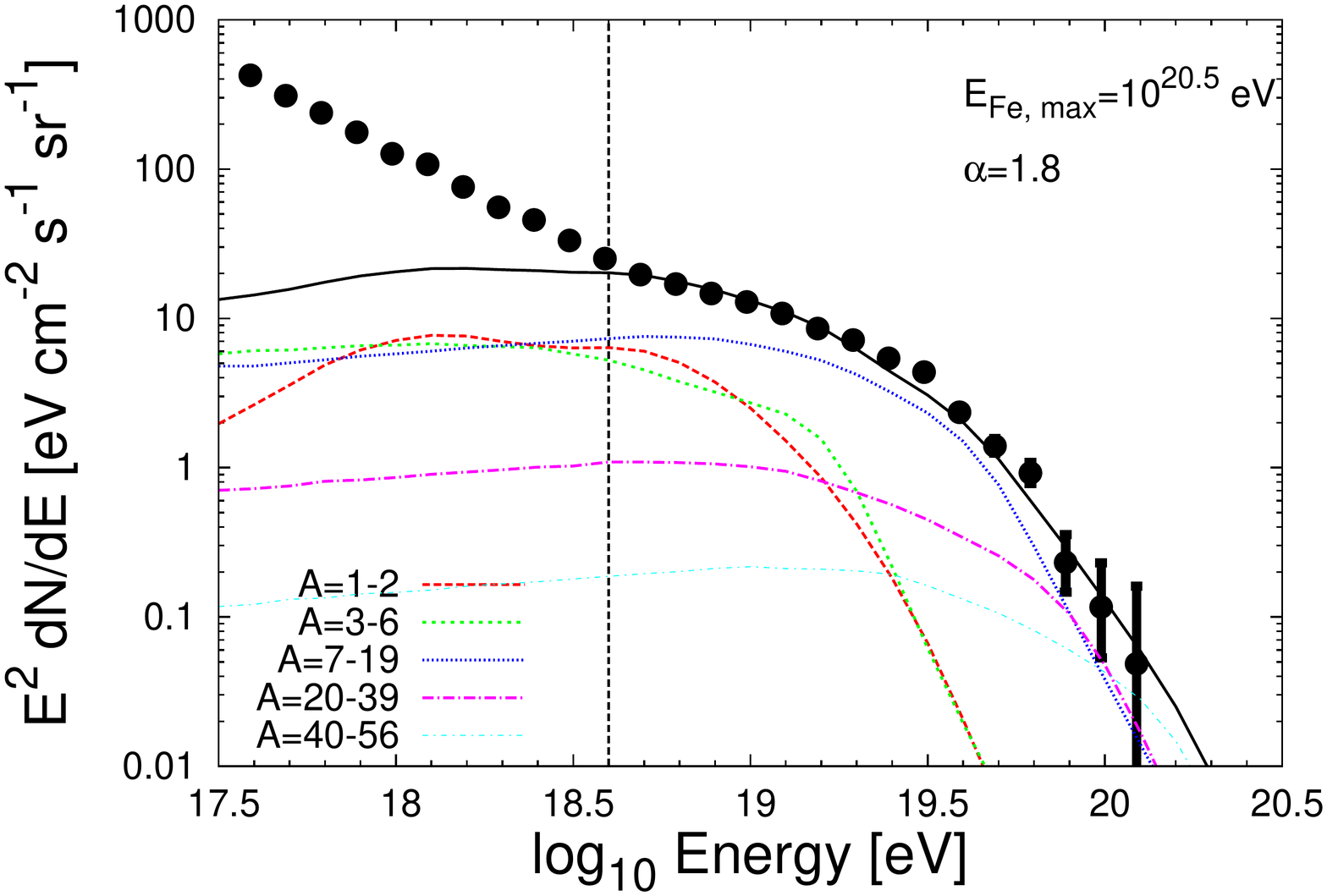}\includegraphics[height=0.25\linewidth, viewport =  0 -32 650 468,clip=true]{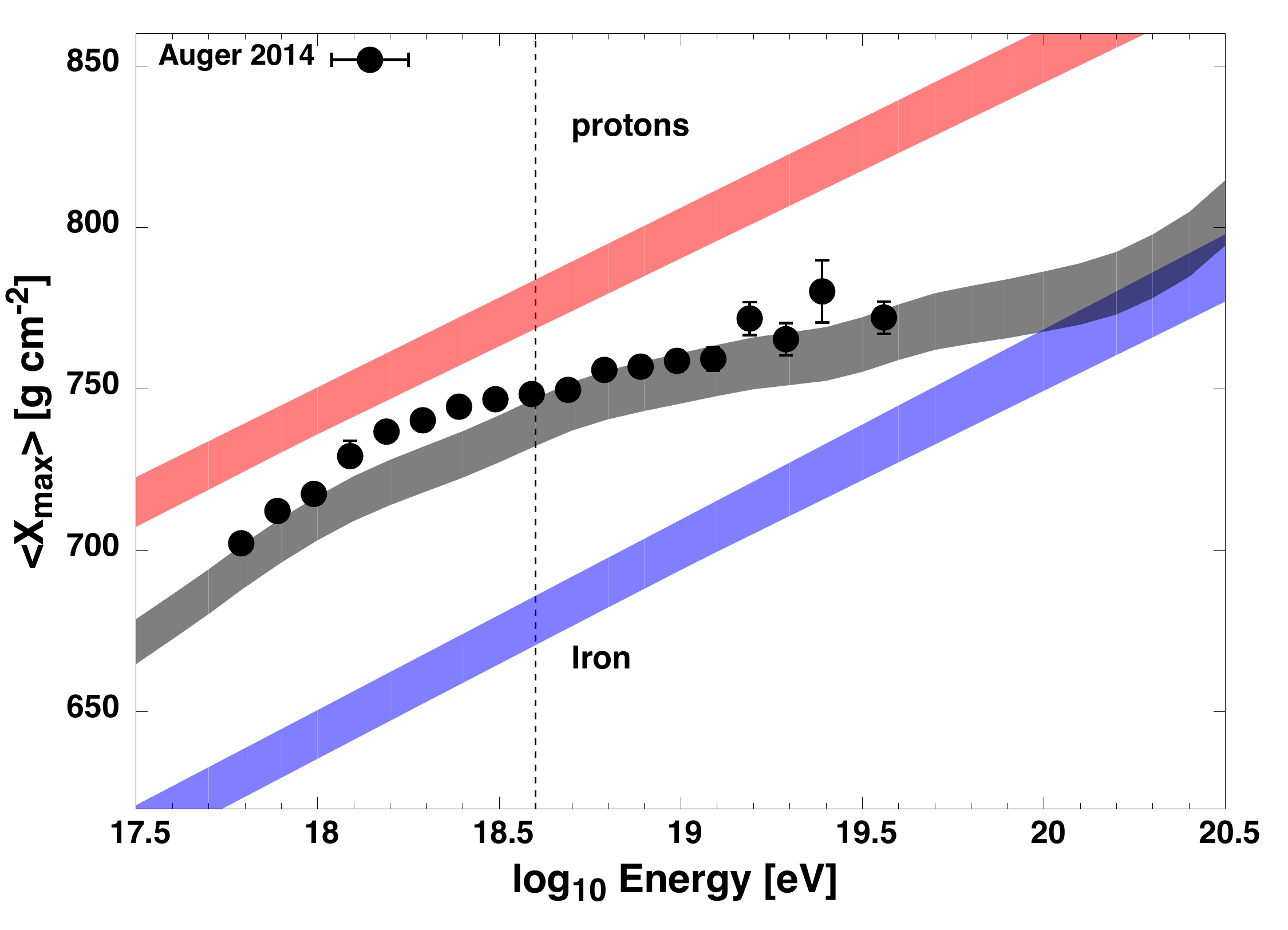}\includegraphics[height=0.25\linewidth, viewport =  0 -32 650 468,clip=true]{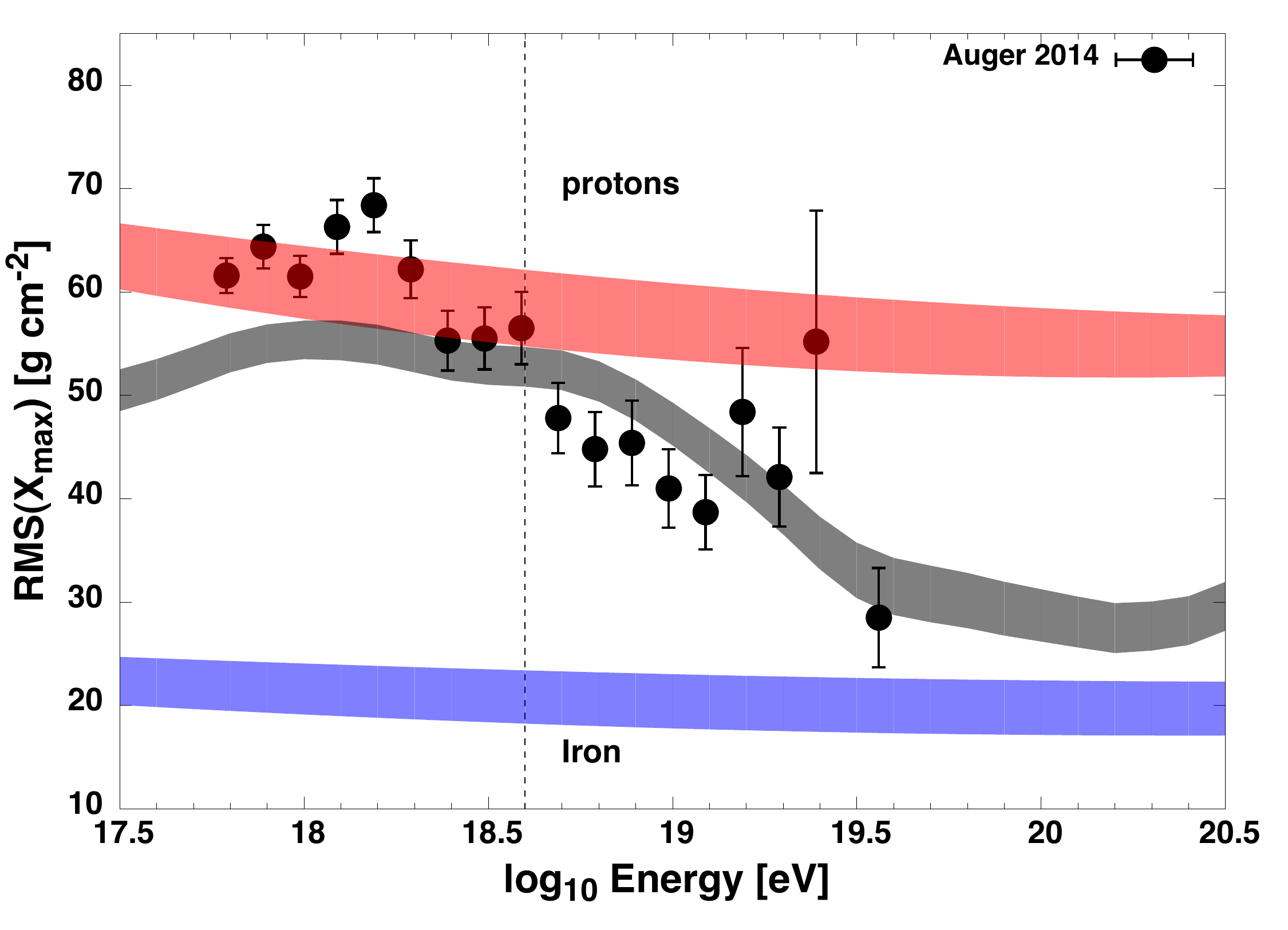}\\
\caption[]{The best-fit models for a source evolution model ${\rm d}N/{\rm d}V_{C} \propto (1+z)^{n}$, up to $z_{\rm max}=3$ with different indices: $n=3$, $n=0$, $n=-3$ and $n=-6$ from top to bottom. In the left frame, we compare the total predicted UHECR spectrum to that measured by the Pierre Auger Observatory. The dashed and dotted curves in this frame denote the contribution from individual nuclear mass groups. In the middle and right frames, we compare the prediction of this model to the depth of shower maximum ($X_{\rm max}$) and its RMS variation, again as measured by Auger~\cite{Aab:2014kda}. 
}
\label{bf_n}
\end{figure}

\section{Fits to the Measured Spectrum and Chemical Composition}
\label{results}

In this section, we describe the statistical methods and main results of our analysis, including the range of UHECR models that provide a good fit to the combined spectrum and composition measurements from the PAO. Given the large number of free parameters being varied in our analysis, we adopt a method that makes use of a Markov Chain Monte Carlo (MCMC) routine, employing the Metropolis-Hastings algorithm to scan through the spectral and composition likelihood landscape and determine the regions that provide a good fit to the data. This method allows local minima regions within the parameter space to be explored
and ``escaped from'' such that the true global minimum can be effectively located. In these scans, the Likelihood function (for given values of $n$ and $z_{\rm max}$) is given by: 
$$
L(f_{\rm p},f_{\rm He},f_{\rm N},f_{\rm Si},E_{\rm Fe, max},\alpha)\propto \exp(-\chi^{2}(f_{\rm p},f_{\rm He},f_{\rm N},f_{\rm Si},E_{\rm Fe, max},\alpha)/2)
,$$ 
where the quantities, $f_i$, denote the fraction of the injected cosmic ray spectrum that consists of the nuclear species, $i$. Note that the parameter $f_{\rm Fe}$ is not free, but is instead fixed by the constraint, $f_{\rm p}+f_{\rm He}+f_{\rm N}+f_{\rm Si}+f_{\rm Fe}=1$. We only include data at energies above $10^{18.6}$~eV in our fit, as it is not clear that the cosmic ray spectrum is dominated by extragalactic particles at lower energies~\cite{2012JCAP...07..031G,Aloisio:2013hya}. Systematic errors in the analysis have been added in quadrature to statistical errors. Throughout our analysis, we adopt the following systematic errors: 14\% on energy, 10~g~cm$^{-2}$ on $\langle X_{\rm max} \rangle$, and 2~g~cm$^{-2}$ on $\mathrm{RMS}(X_{\rm max})$~\cite{Aab:2014kda}.

Figure~\ref{bf_n} shows the best-fit spectra, $\langle X_{\rm max} \rangle$ and $\mathrm{RMS}(X_{\rm max})$ distributions for four different evolution indicies: $n=3$, $0$, $-3$, and $-6$.
We begin by considering the case with no source evolution ($n=0$). In this case, the best-fit is found for sources which inject a large fraction of helium ($f_{\rm He} = 0.53$) and particles in the nitrogen group ($f_{\rm N} = 0.29$), along with smaller but non-negligible quantities of protons ($f_p=0.17$) and very few heavier nuclei ($f_{\rm Si}=0.0$, $f_{\rm Fe}=0.01$). As mentioned in the introduction, the injected spectrum is very hard in this model, with an index of $\alpha=1.1$ and $E_{\rm Fe, max}=10^{20.2}$ eV. This is in considerable contrast to the softer spectra generally predicted by Fermi acceleration, $\alpha \simeq 2$. The results for the best-fit model with no source evolution are shown in Fig.~\ref{bf_n}.

We also find good fits for positive ($n=3$) and negative ($n=-3$ \& $-6$) source evolution as shown in the other panels of Fig.~\ref{bf_n}, in each case favoring models with large fractions of helium and nitrogen at injection and $E_{\rm Fe, max} \simeq (1.5-3) \times 10^{20}$~eV (see Table~\ref{bestfittable}). The main difference among the best-fit models is the value of the injected spectral index, which varies from $\alpha=$0.6 for $n=3$ to $\alpha=$1.8 for $n=-6$. In Fig.~\ref{bf_fractions} we compare the best-fit models for the four different evolution models to the mass composition inferred by PAO via the template-based fit of the $X_{\rm max}$ distribution~\cite{Aab:2014aea}. The vertical dashed line indicates the low energy threshold of our fit. For all evolution models the observed best-fit mass spectrum after propagation agrees with the PAO observation within uncertainties. In particular, heavy mass groups beyond nitrogen have only a small contribution in the observed spectrum.

\begin{figure}[t]
\centering\leavevmode
\includegraphics[width=0.48\linewidth, angle=0]{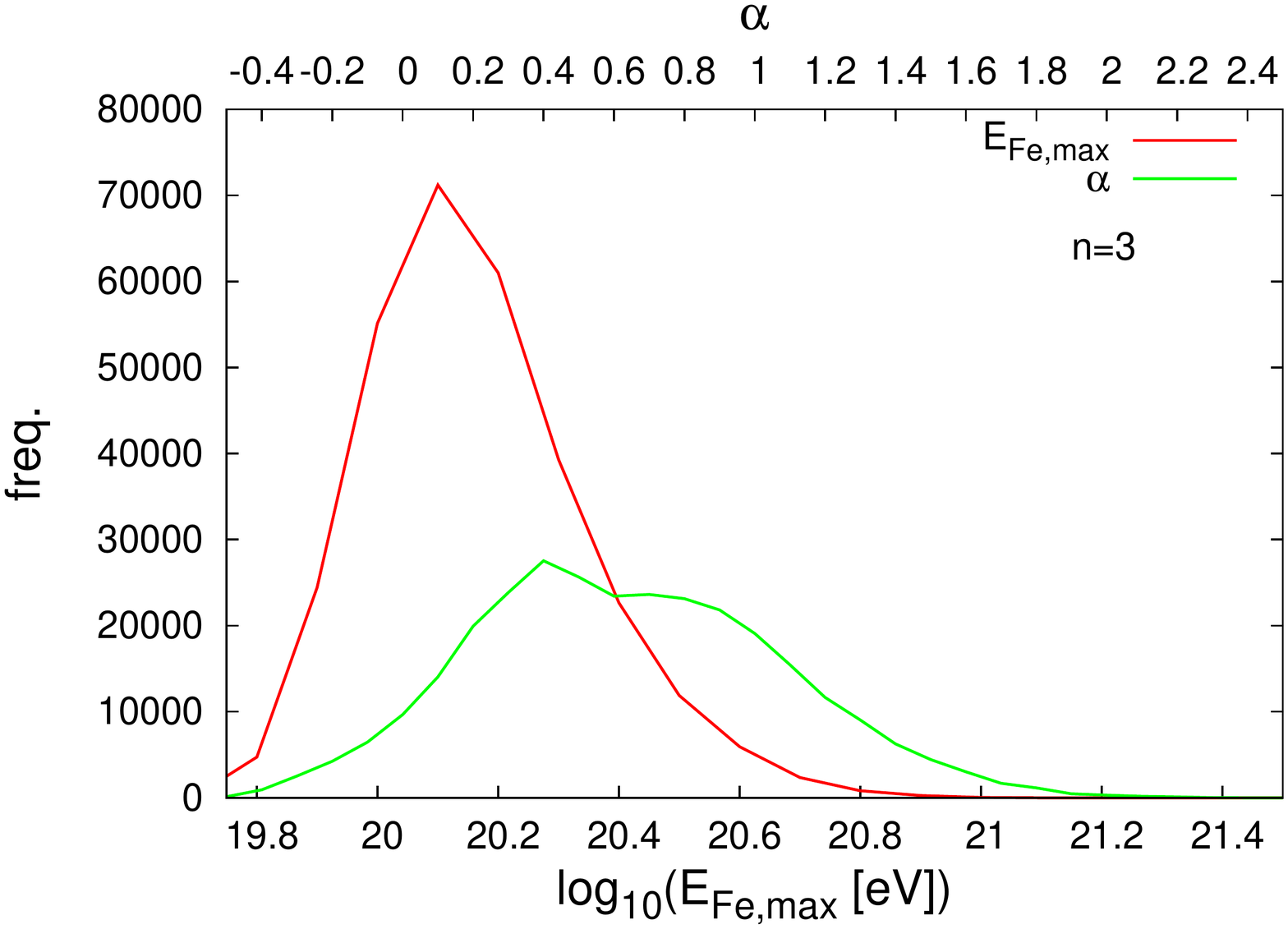}
\hspace{-0.9cm}
\includegraphics[width=0.48\linewidth, angle=0]{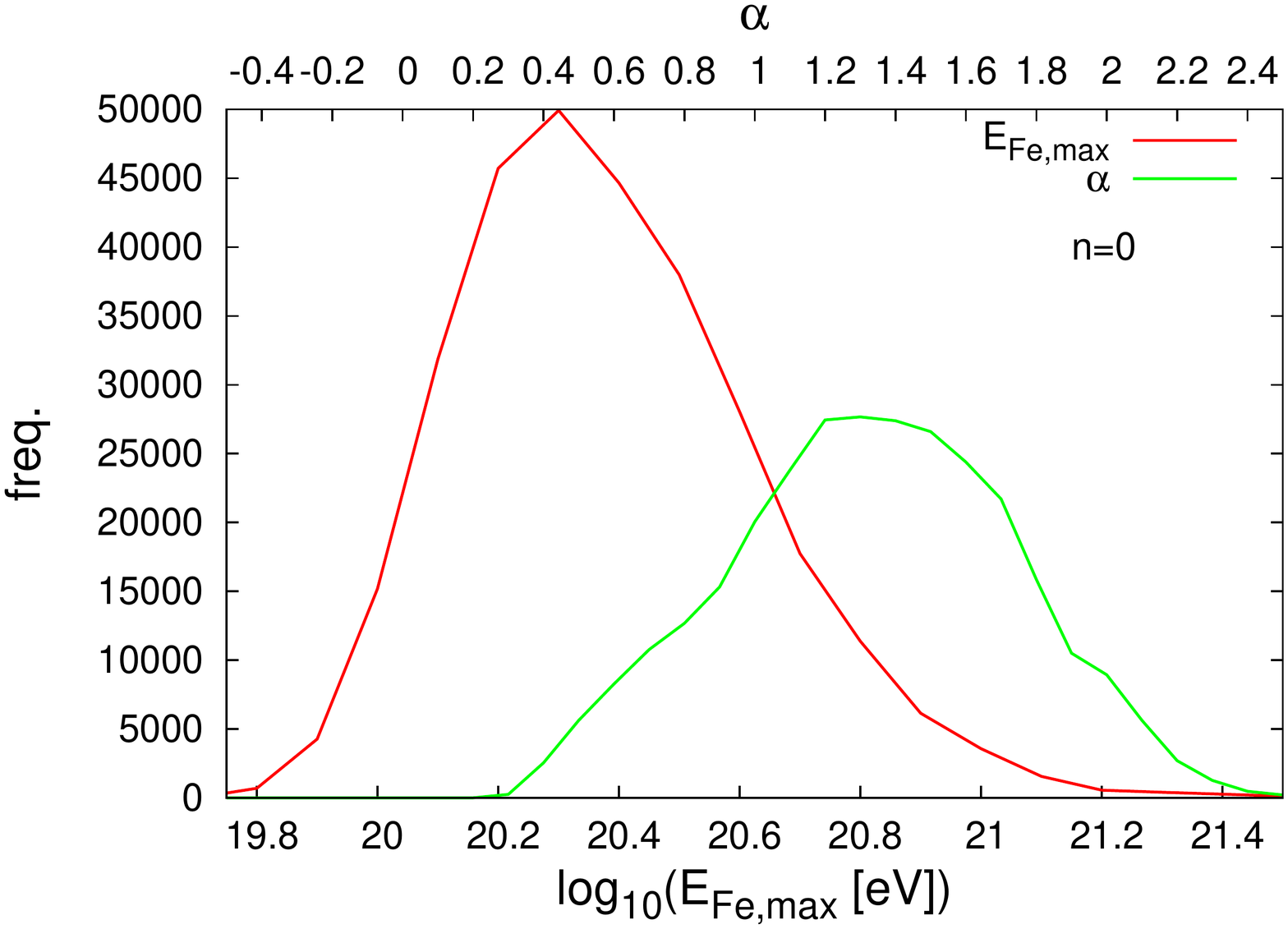}\\
\vspace{-0.9cm}
\includegraphics[width=0.48\linewidth, angle=0]{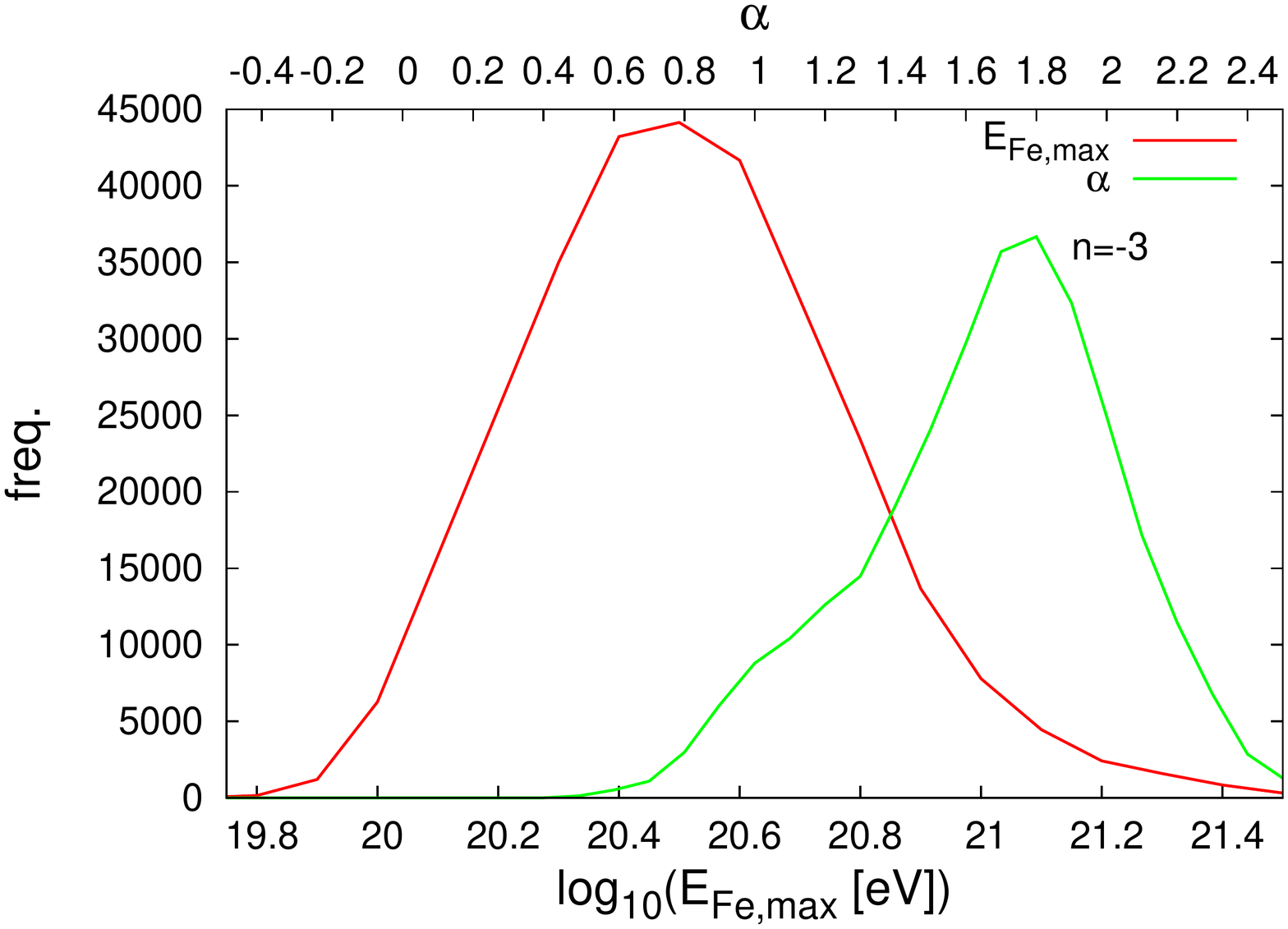}
\hspace{-0.9cm}
\includegraphics[width=0.48\linewidth, angle=0]{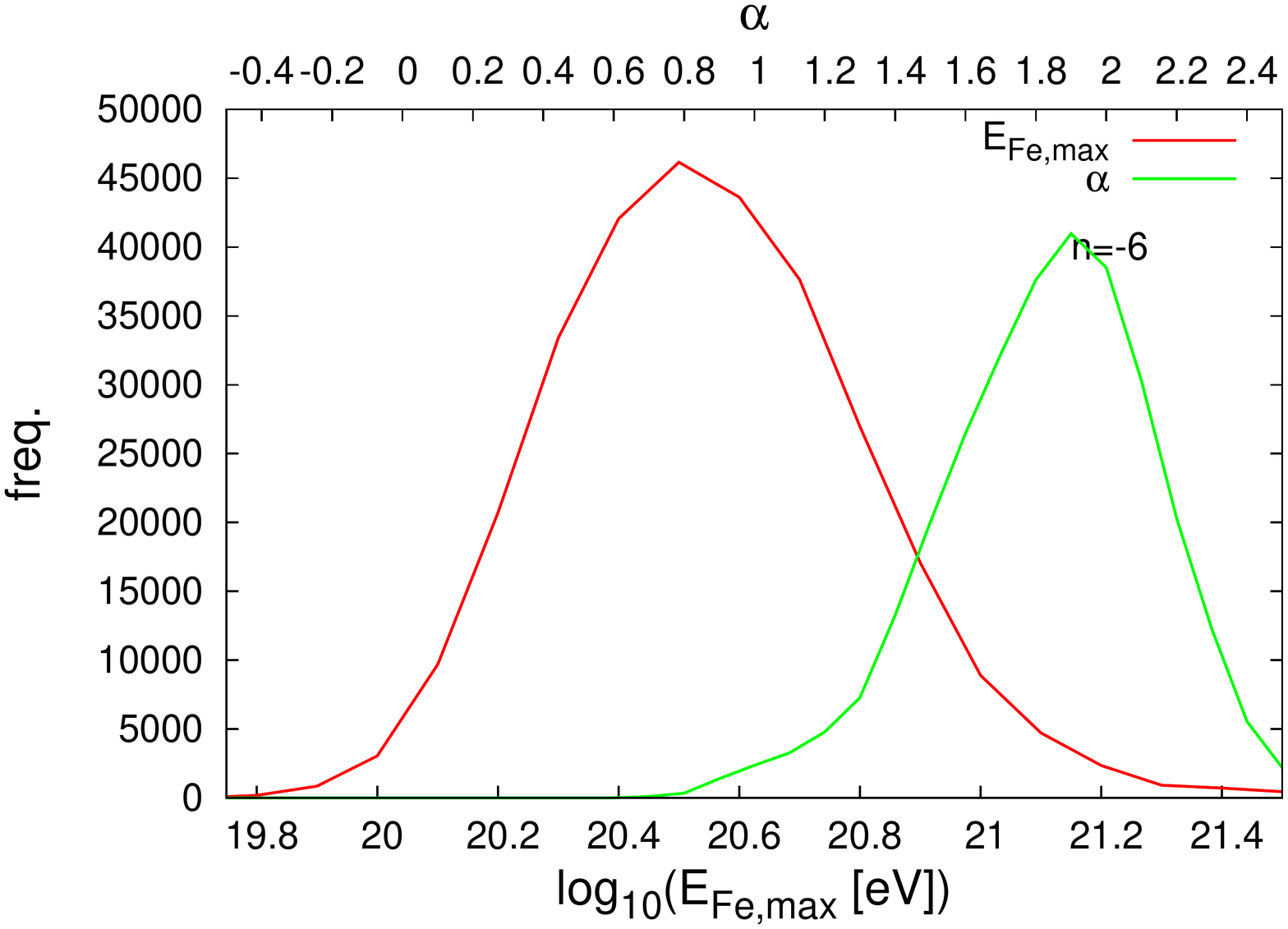}
\vspace{-0.6cm}
\caption{Frequency plots from the MCMC scan of the spectral parameters $\alpha$ and $E_{\rm Fe,max}$ for four values of the source evolution parameter, $n$. Models with no evolution ($n=0$) or positive evolution ($n=3$) favor very hard spectral indices for the injected spectrum, whereas models with negative evolution ($n=-3$ \& $n=-6$) prefer significantly softer injected spectra.}
\label{const}
\end{figure}

\begin{figure}[t]\centering
\includegraphics[width=0.65\linewidth, angle=0]{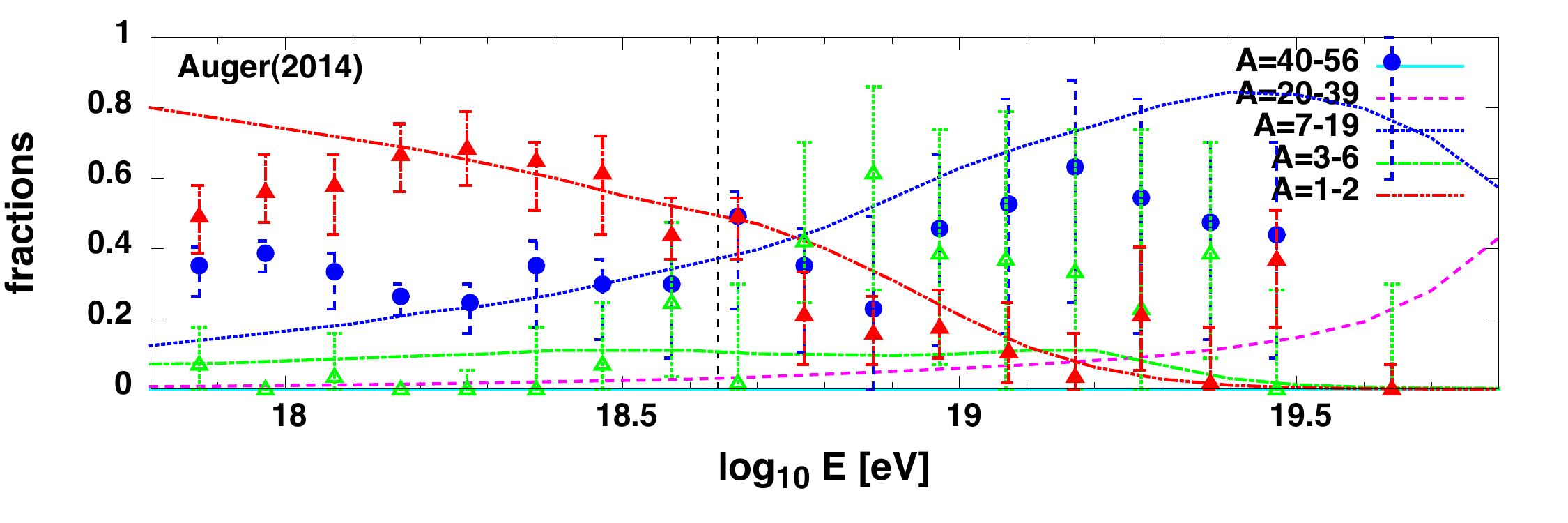}\\
\includegraphics[width=0.65\linewidth, angle=0]{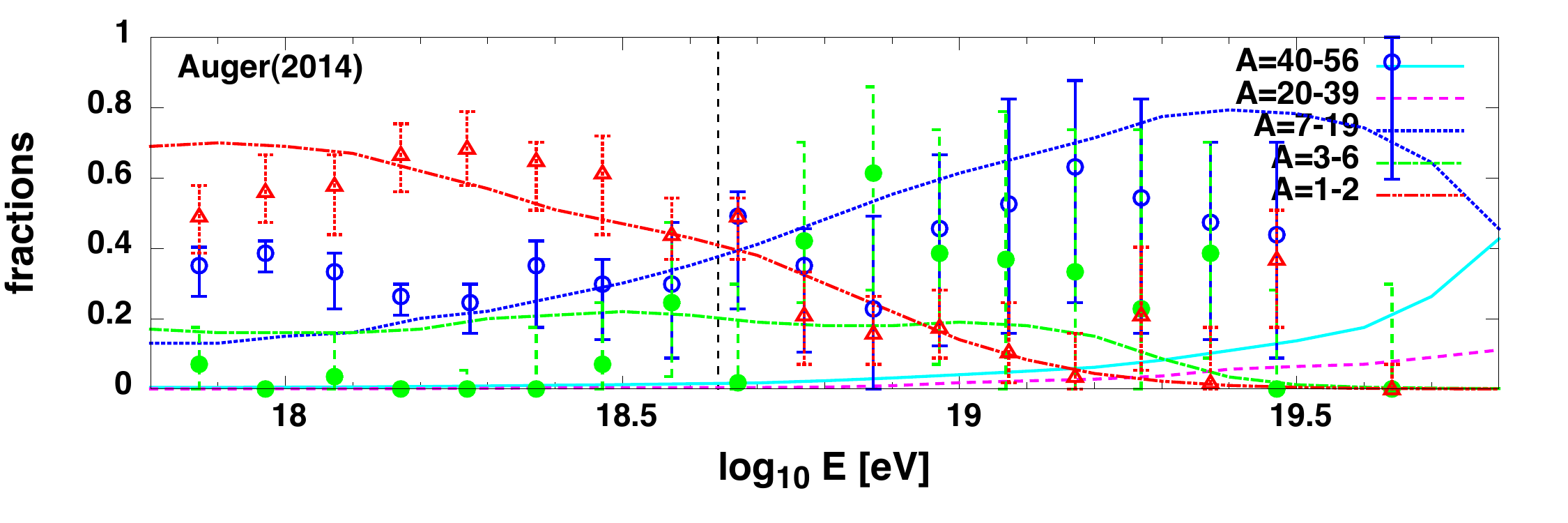}\\
\includegraphics[width=0.65\linewidth, angle=0]{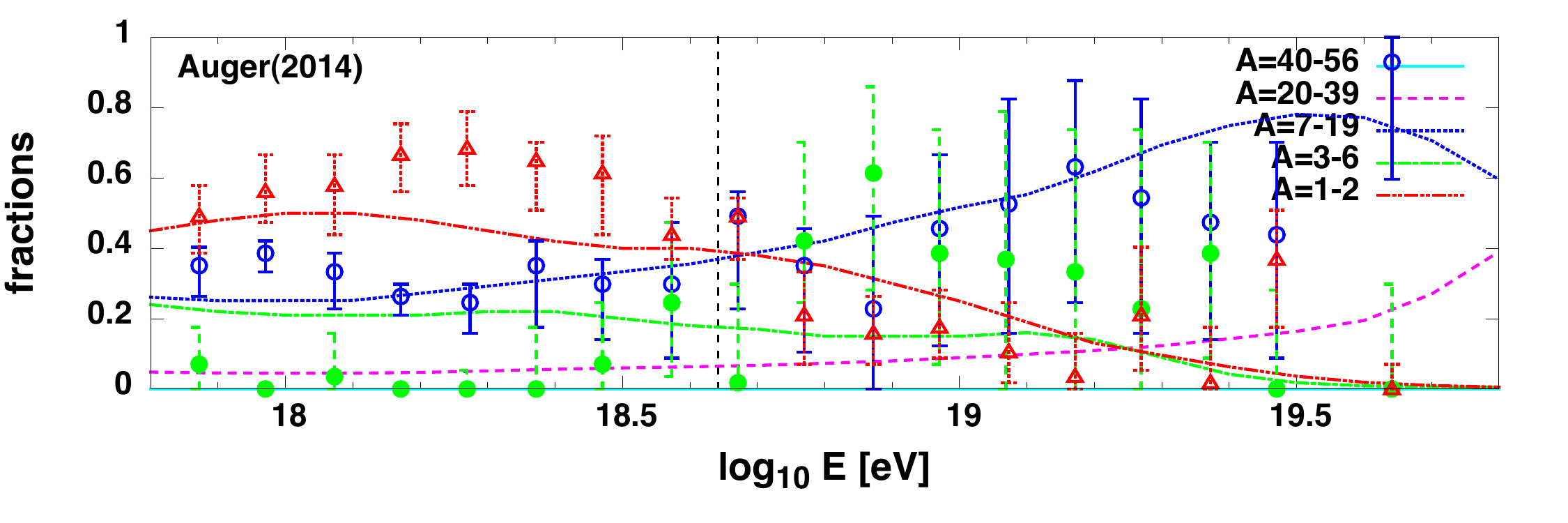}\\
\includegraphics[width=0.65\linewidth, angle=0]{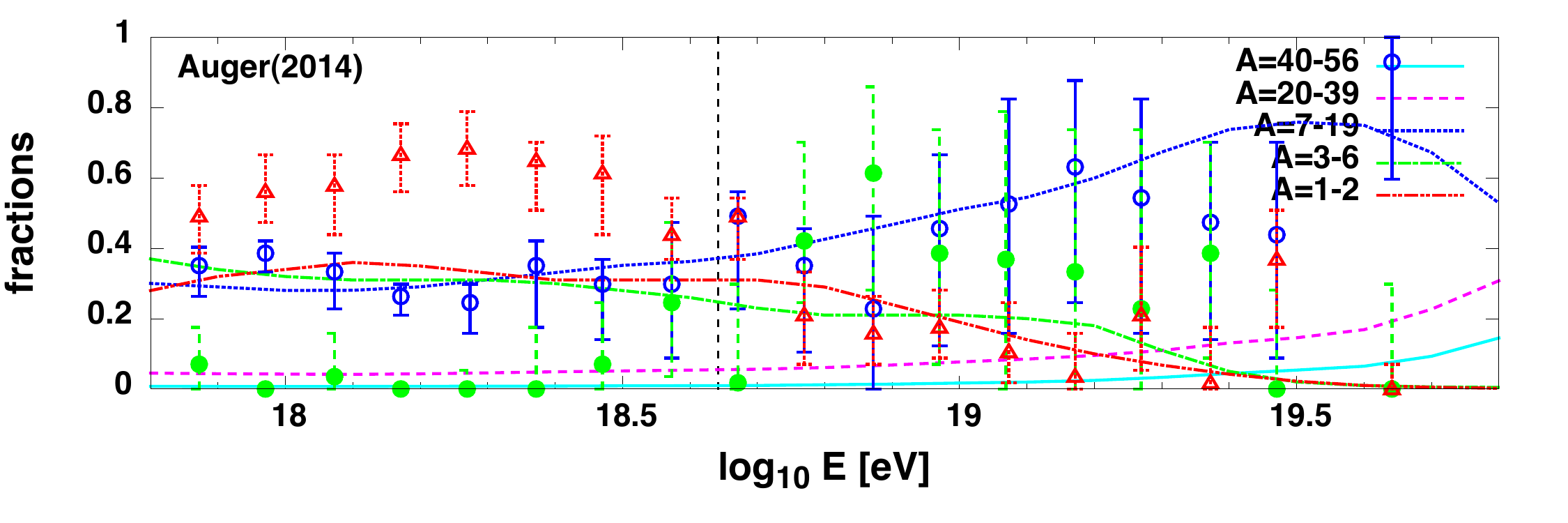}
\caption[]{The predicted chemical composition of the best-fit UHECR spectrum at Earth compared to that reported by the Auger Collaboration~\cite{Aab:2014aea}. Results are shown for a source evolution model ${\rm d}N/{\rm d}V_{C} \propto (1+z)^{n}$, up to $z_{\rm max}=3$ with different indices: $n=3$, $n=0$, $n=-3$ and $n=-6$ from top to bottom. For these results the EPOS-LHC~\cite{Pierog:2013ria} hadronic model is adopted and the data and model line colors indicate the nuclear species range applicable.}
\label{bf_fractions}
\end{figure}

Moving beyond best-fit models, we show in Fig.~\ref{const} the weighted distributions of the spectral parameters $\alpha$ and $E_{{\rm Fe, max}}$ that are found by our MCMC scan, for each evolutionary model. Although a modest shift in $E_{\rm Fe, max}$ is observed with $n$, the greatest variation is observed among the values of $\alpha$ that are favored. Models with no evolution or positive evolution favor very hard spectral indices for the injected spectrum, whereas models with negative evolution prefer softer injected spectra. In Table~\ref{bestfittable}, we also provide the posterior mean and standard deviation for these parameters, as well as for the injected composition fractions.

In addition to the spectral parameterization described in Eq.~\ref{source_spec}, we have also considered spectra with sharper than exponential cutoffs. Such spectra are motivated from particle acceleration scenarios in which electrons are both accelerated and radiatively cooled within the source environment~\cite{Zirakashvili:2006pv,Stawarz:2008sp}. No qualitative changes to our conclusions were found in this case, however.

It should be noted that the distribution of the spectral parameters presented in Fig.~\ref{const} are correlated, and are thus not independently constrained. As pointed out previously in Ref.~\cite{Taylor:2013gga} (which considered the specific case of $n=3$), the models with the largest values of $\alpha$ also exhibit large values of $E_{\rm Fe, max}$. Since the time of that study, the systematic uncertainties associated with the PAO's measurements have been reduced considerably, and updates from recent LHC runs have reduced in the variation between the predictions of various hadronic models. After taking these improvements into account, it is apparent that the scenarios considered in Ref.~\cite{Taylor:2013gga}, which were previously found to be in reasonable agreement with the data, are now disfavored.

We also note that the proton to helium ratios in Table~\ref{bestfittable}, within their 1~sigma standard deviation regions, appear unphysically small. This feature appears to be an artifact of the energy threshold of $10^{18.6}$~eV adopted. Indeed, a reduction of this energy threshold in the analysis allows much larger ratios. Such a decrease of the threshold energy, however, potentially also demands a more complicated setup. With the composition of the flux in the energy range below $10^{18.6}$~eV and above $10^{18}$~eV being indicated to be light, we highlight these complications through the consideration of an additional proton flux component, whose spectral index/source evolution is not tied to that of the heavier nuclei, to account for this flux. We find that for reasonable fits to the spectrum to be obtained, either softer spectral indices than for the $>10^{18.6}$~eV component (as found previously by others \cite{Aloisio:2013hya}), or a stronger source evolution are required for this additional light component.

\begin{table}[t]\renewcommand*{\arraystretch}{1.5}
\begin{tabular}{c|cc|cc|cc|cc}
\toprule
 & \multicolumn{2}{c|}{$n=-6$} & \multicolumn{2}{c|}{$n=-3$} &\multicolumn{2}{c|}{$n=0$} &\multicolumn{2}{c}{$n=3$}  \\
Parameter & \mini{1.2cm}{\scriptsize Best-fit\\ Value }& \mini{2.4cm}{\scriptsize Posterior Mean \&\\ Standard Deviation} & \mini{1.2cm}{\scriptsize Best-fit\\ Value }& \mini{2.4cm}{\scriptsize Posterior Mean \&\\ Standard Deviation}& \mini{1.2cm}{\scriptsize Best-fit\\ Value }& \mini{2.4cm}{\scriptsize Posterior Mean \&\\ Standard Deviation}& \mini{1.2cm}{\scriptsize Best-fit\\ Value }& \mini{2.4cm}{\scriptsize Posterior Mean \&\\ Standard Deviation}\\[0.3cm]
\hline
$f_{p}$ & 
$0.03$ & $0.14\pm0.12$ & 
$0.08$& $0.15\pm0.13$ & 
$0.17$& $0.17\pm0.16$ & 
$0.19$& $0.20\pm0.16$ \\
$f_{\rm He}$ & 
$0.50$ & $0.21\pm0.17$ & 
$0.42$& $0.17\pm0.16$ & 
$0.53$& $0.20\pm0.17$ & 
$0.32$& $0.23\pm0.20$ \\ 
$f_{\rm N}$ & 
$0.40$ & $0.50\pm0.18$ &
$0.42$ & $0.51\pm0.19$ &
$0.29$ & $0.47\pm0.19$ & 
$0.43$ & $0.45\pm0.21$ \\
$f_{\rm Si}$ &
$0.06$ & $0.11\pm0.12$ &
$0.08$ & $0.12\pm0.13$ &
$0.0$ & $0.11\pm0.12$ &
$0.06$ & $0.078\pm0.086$ \\
$f_{\rm Fe}$  & 
$0.01$ & $0.052\pm0.039$ &
$0.0$ & $0.053\pm0.042$ &
$0.01$ & $0.050\pm0.038$ &
$0.0$ & $0.044\pm0.034$ \\
$\alpha$ & 
$1.8$ & $1.83\pm0.31$ &
$1.6$ & $1.67\pm0.36$ &
$1.1$ & $1.33\pm0.41$ &
$0.6$ & $0.64\pm0.44$ \\
$\log_{10}\!\!\left( \frac{E_{\rm Fe, max}}{\rm eV}\right)$ &
$20.5$ & $20.55\pm0.26$ &
$20.5$ & $20.52\pm0.27$ &
$20.2$ & $20.38\pm0.25$ &
$20.2$ & $20.16\pm0.18$ \\[0.1cm]
\toprule
\end{tabular}
\caption[]{The parameters which provide the best-fit to the spectrum, $\langle X_{\rm max} \rangle$, and RMS$(X_{\rm max})$, as measured by the Pierre Auger Observatory, for four choices of the source evolution model. Models with no evolution ($n=0$) or positive evolution ($n=3$) favor very hard spectral indices for the injected spectrum, whereas models with negative evolution ($n=-3$, -6) prefer significantly softer injected spectra. We also show the posterior mean value and standard variation of the injected spectrum and composition parameters resulting from our MCMC scan.}
\label{bestfittable}
\end{table}

\section{Constraints from Cosmogenic Gamma-Rays and Neutrinos}
\label{constraints}

In the previous section, we compared the predictions of a wide range of UHECR models (scanned by MCMC) to the spectrum and composition measurements reported by the Pierre Auger Collaboration. In some respects, these results were found to be largely insensitive to variations in the source evolution parameter, $n$. In particular, we find good fits to the combined spectrum and composition data for each evolution scenario, and the injected chemical composition preferred by the fit is only mildly impacted by our choice of the evolution model. In other ways, however, our results depend very strongly on the choice of source evolution.  Most notably, the injected spectral index required to fit the spectrum measured by the PAO varies considerably with evolution, favoring $\alpha \lsim 1$ for models with significant positive evolution and $\alpha \simeq 2$ for models with significant negative evolution.

In addition to the impact on the spectral index the flux of cosmogenic gamma-rays and neutrinos from the propagation of UHECR varies considerably with the evolutionary model being considered. Interactions of the UHECRs with the cosmic microwave and infrared backgrounds invariably generate a population of high-energy neutrinos, known as the cosmogenic neutrino flux. In particular, for proton-dominated models, photo-pion production with the CMB leads to a cutoff in the spectrum at $E_{\rm GZK} \simeq 50$~EeV~\cite{Greisen:1966jv,Zatsepin:1966jv,Berezinsky:1969qj}. The resulting neutrino flux becomes maximal at a few EeV~\cite{Beresinsky:1969qj}.

If the UHECR spectrum is dominated by intermediate mass nuclei, as indicated by cosmic ray observations by the Auger Collaboration, however, the resonant interaction of cosmic ray nucleons with the CMB is shifted to higher energies, ($A/56$) $\times~3$~ZeV, where $A$ is the atomic mass. For heavy or intermediate mass nuclei, the energy at which neutrinos are most efficiently generated is shifted upward, to where significantly fewer cosmic rays exist.  As a result, nuclei-dominated cosmic ray models predict significantly lower fluxes of cosmogenic neutrinos than proton-dominated scenarios. 

Due to this increased threshold of cosmogenic neutrino production for cosmic ray nuclei, cosmic radiation at infrared, optical, and ultra-violet wavelengths can be important targets, and are included in most modern cosmogenic neutrino calculations~\cite{Hooper:2004jc,Ave:2004uj,Hooper:2006tn,Allard:2006mv,Anchordoqui:2007fi,Aloisio:2009sj,Kotera:2010yn,Decerprit:2011qe,Ahlers:2011sd}. In general, the peak of the cosmogenic neutrino spectrum predicted for nuclei-dominated models is shifted downward, to the 1-100 PeV range, and the overall normalization is reduced relative to proton-dominated models, generally below present experimental sensitivities. The overall cosmogenic neutrino flux also depends on the maximal energies injected and the redshift evolution of UHECR sources. An estimate of the lower limit of these pessimistic models was given in Ref.~\cite{Ahlers:2012rz}.

\begin{figure}[t]
\centering\leavevmode
\includegraphics[width=0.49\linewidth,viewport= 10 0 520 400,clip=true]{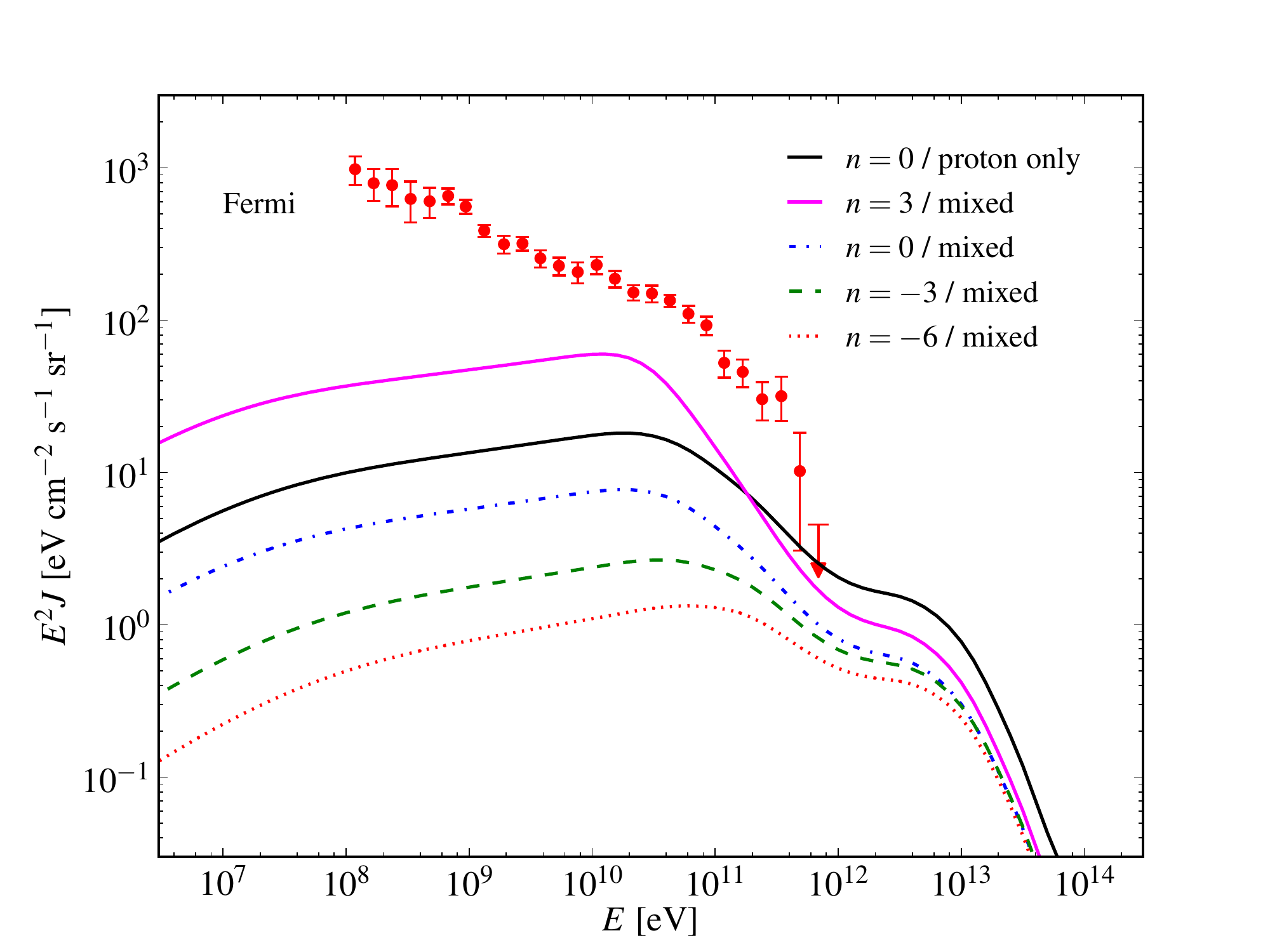}\hfill\includegraphics[width=0.49\linewidth,viewport= 10 0 520 400,clip=true]{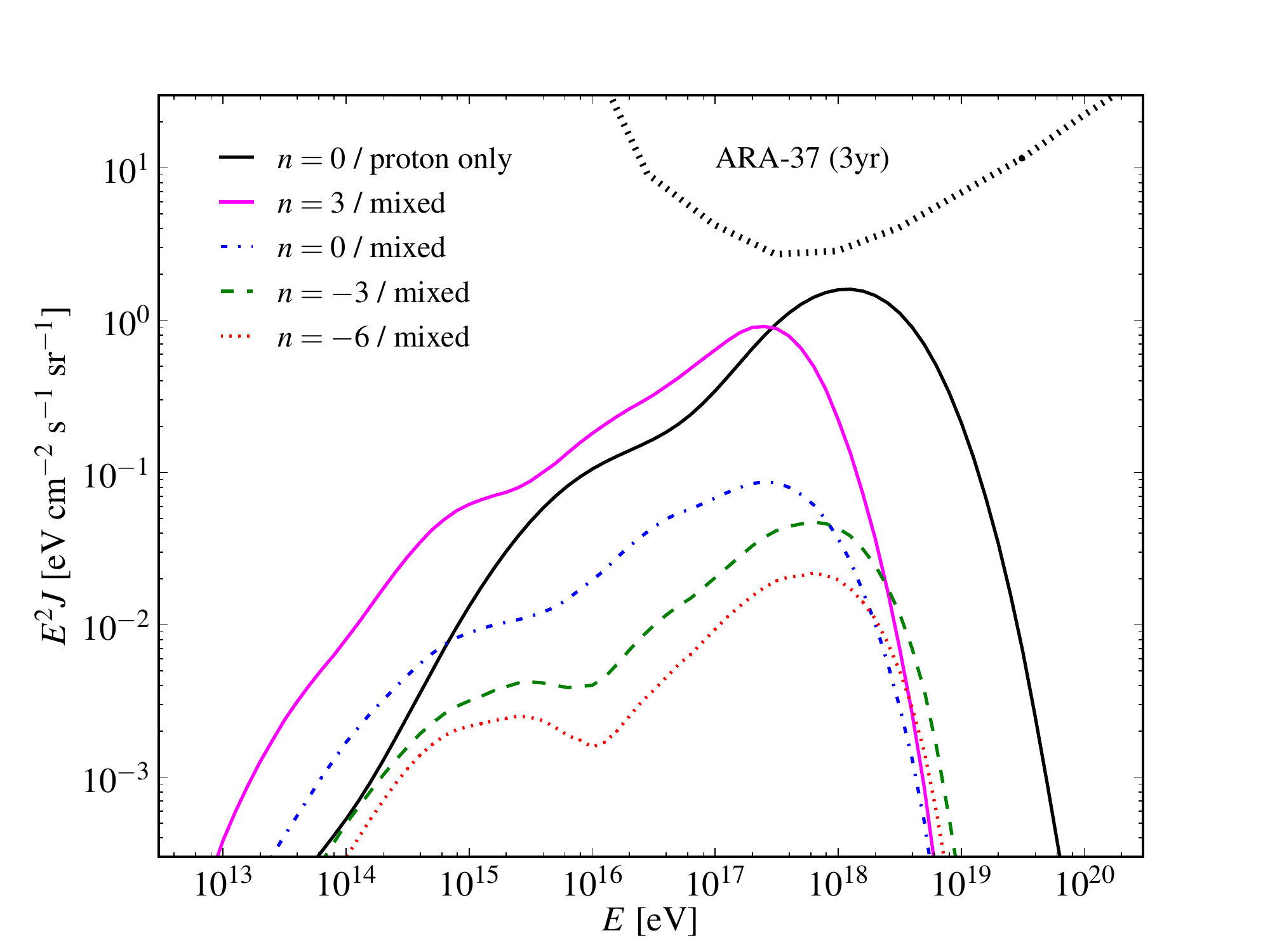}
\caption{The contribution of cosmogenic gamma-rays (left plot) and cosmogenic neutrinos (all flavor; right plot) from UHECR propagation for the best-fit models in four different source evolution scenarios ($n=3$, $0$, $-3$, $-6$). For comparison, we also show the contribution resulting from a purely-proton model with no source evolution ($\alpha=2.05$, $E_{{\rm max}, p}=10^{19.6}$~eV, $n=0$). The left panel shows the isotropic gamma-ray background inferred by {\it Fermi}~\cite{Ackermann:2014usa}. All best-fit models considered here are consistent with the gamma-ray background. In the models favored by the data (nuclei-dominated, $n\lsim 0$), also very low cosmogenic neutrino fluxes are predicted. The thick dotted line in the right panel shows the sensitivity of the proposed ARA-37 after three years of observation~\cite{Allison:2014kha}.}
\label{secondaries}
\end{figure}

In the right panel of Fig.~\ref{secondaries}, we show the cosmogenic neutrino flux (summed over all flavors) for the best-fit models for four choices of redshift evolution ($n=3$, $0$, $-3$, $-6$).  The higher energy peak at EeV energies is due to neutrinos produced on the CMB photons, whereas the lower peak at PeV energies is largely due to interactions with the infrared background~\cite{Franceschini:2008tp}. These predictions are well below the present limits placed by the IceCube~\cite{Aartsen:2013dsm}, Anita~\cite{Gorham:2010kv} and PAO~\cite{Abreu:2012zz} Collaborations, and are even too low to be reached by future radio Cherenkov observatories, such as ARA~\cite{Allison:2014kha}, or ARIANNA~\cite{Barwick:2014pca}. For orientation we show the projected ARA-37 3yr sensitivity in the plot. These predicted cosmogenic neutrino fluxes are also much lower than the extraterrestrial neutrino flux reported by the IceCube Collaboration, which is near the level of $E_\nu^2J^{\rm IC}_{\nu_\alpha} \simeq (9.5\pm3.0)~{\rm eV} \,{\rm s}^{-1}\, {\rm cm}^{2} \,{\rm sr}^{-1}$ over the energy range of $\sim$10 TeV-PeV~\cite{Aartsen:2014gkd}. It has been speculated that this emission could be indirectly related to the flux of UHECRs via calorimetric processes in cosmic ray reservoirs affecting the low energy tail of the emission spectrum~\cite{Katz:2013ooa,Cholis:2012kq}. Such environments have been discussed within the context of starburst galaxies~\cite{Loeb:2006tw,Tamborra:2014xia}, Galactic outflows \cite{Feldmann:2012rx,Taylor:2014hya}, or clusters of galaxies~\cite{Berezinsky:1996wx,Murase:2008yt,Zandanel:2014pva}.

The photo-hadronic interactions with the cosmic radiation backgrounds responsible for the flux of cosmogenic neutrinos also generate a flux of high-energy gamma-rays. In addition, non-resonant Bethe-Heitler pair production via cosmic ray scattering off CMB photons creates a population of high energy electrons and positrons. However, these electro-magnetic contributions undergo rapid cascades via pair production and inverse-Compton scattering in the CMB, resulting in a cascaded sub-TeV gamma-ray spectrum. These gamma-ray contributions are shown for the best-fit models in the left plot of Fig.~\ref{secondaries} in comparison to the isotropic gamma-ray background inferred by {\it Fermi}~\cite{Ackermann:2014usa}.

For the best-fit model with no source evolution ($n=0$, $z_{\rm max}=3$), the electromagnetic cascades resulting from UHECRs produce approximately 10\% of the measured isotropic gamma-ray background above $10^{11}$~eV~\cite{Ackermann:2014usa}. In models with positive evolution ($n=3$, $z_{\rm max}=3$), the contribution increases to 20\%, while models with negative evolution contribute significantly less to the diffuse gamma-ray flux. This result is not surprising, as a reduction in the evolution parameter amounts to moving the sources of the UHECRs to more local distances, reducing their mean propagation time through the extragalactic radiation fields that give rise to these losses (predominantly the CMB). The entirety of the isotropic gamma-ray background observed by {\it Fermi} can be easily accommodated by the sum of known gamma-ray source classes (blazars, starforming galaxies, and radio galaxies)~\cite{Ajello:2015mfa,Cholis:2013ena,Inoue:2012cs}, and {\it Fermi}'s most recent measurement leaves relatively little room for an additional component originating from UHECR propagation (see, for example, Fig.~3 of Ref.~\cite{Ajello:2015mfa}). However, a contribution at the level of $\sim$10-20\% at energies above $10^{11}$~eV is likely compatible with the current {\it Fermi} data.

We also briefly consider the cascade contribution from a separate dominant extragalactic proton component in the range $10^{18}$--$10^{18.6}$~eV for different source evolution models. We find that such a setup gives rise to a factor of $\sim$2--4 increase in the cascade flux levels, over our previous results, for the different evolution scenarios ($n=3,-6$) considered. The increase by a factor of $\sim 2$ was found for the $n=-6$ evolution and the increase by a factor of $\sim 4$ for the $n=3$ evolution, with intermediate evolution scenarios receiving intermediate flux increases. Though such increases still sit below the isotropic background level, this result highlights the importance in determining the transition energy at which the extragalactic sources start to dominate the cosmic ray spectrum, as already highlighted by others \cite{Ahlers:2010fw,Berezinsky:2010xa}.

We also note that the model-to-model variation in the contribution to the isotropic gamma-ray background predicted by our MCMC scan is quite small (when the source evolution model is fixed). For the four source evolution models considered, we find that the predicted contribution to the gamma-ray background is always less than 20\%.

\section{Summary and Discussion}
\label{conclusion}

In this study, we have used a Markov Chain Monte Carlo approach to explore a wide range of models for the sources of the highest energy cosmic rays, varying the injected spectra, chemical composition, and redshift distribution. After comparing the predictions of these models to the most recent data from the Pierre Auger Observatory~\cite{Aab:2014kda}, we find that models dominated by intermediate mass nuclei are strongly favored. This is consistent with a template-based composition analysis of the $X_{\rm max}$ distribution~\cite{Aab:2014aea}. Among these models, those without redshift evolution (corresponding to a constant number of sources per comoving volume, $n=0$) are capable of fitting the observed spectral shape only if the injected spectrum is very hard, with a spectral index of $\alpha\simeq1.3$.  If the number of sources increases at low redshifts ($n<0$), however, softer spectra consistent with the expectations of Fermi acceleration ($\alpha \simeq 2$) are favored. 

We argue that these considerations significantly favor UHECR models with negative source evolution, and strongly disfavor positive evolution scenarios. These findings both strengthen previous results indicating a local proximity of UHECR sources \cite{Taylor:2011ta}, and further suggests a reduction in the source comoving density at higher redshifts.


We also show that the contribution of electro-magnetic cascades from the production of high-energy gamma-rays, electrons and positrons during the propagation of UHE CRs are consistent with the isotropic gamma-ray background inferred by the {\it Fermi} Gamma-Ray Space Telescope~\cite{Ackermann:2014usa}. The flux of cosmogenic neutrinos is also consistent with upper limits. In general, a strong negative evolution, {\it i.e.}~a stronger contributions of local sources decreases the flux of these contribution and makes a direct observation more challenging.

In light of these results, it is interesting to contemplate which classes of UHECR sources could plausibly be distributed preferentially within the low-redshift universe. One interesting source class are low-luminosity gamma-ray BL Lacertae (BL Lac) objects. While {\it Fermi} measurements have revealed that the number density of bright BL Lacs peaks at a fairly high redshift of $z\simeq 1.2$, the more numerous low-luminosity ($L_{\gamma}<10^{44}$~erg~s$^{-1}$) and high-synchrotron peaked members of this population exhibit negative source evolution, and thus are overwhelmingly distributed at low redshifts~\cite{Ajello:2013lka}.  And while many other possibilities remain, the results of this paper provide some support for this class of objects as the sources of the UHECRs. Given the modest number of low-redshift BL Lacs observed by {\it Fermi}, the lack of anisotropy in the UHECR spectrum would likely require significant deflection by magnetic fields, whether throughout the intergalactic medium, or within the volume of local group or Milky Way. 

Lastly, the contribution of an additional extragalactic proton component, to explain the flux below $10^{18.6}$~eV and above $10^{18}$~eV, was also considered. The composition of this component, indicated to be light, motivating the consideration for it to be extragalactic in origin. Curiously, however, this component was found to require either a softer source spectral index, as also noted by others previously \cite{Aloisio:2013hya}, or a stronger source evolution parameter to the higher energy component we have focused on. The additional gamma-ray cascade contributions from this component was found to be a factor of $\sim$2--4 larger than that from the $>10^{18.6}$~eV contributions considered. This result highlights the importance in the determination of the transition energy at which extragalactic sources start to dominate the cosmic ray flux.

\bigskip

{\bf Acknowledgements}
DH is supported by the US Department of Energy under contract DE-FG02-13ER41958. Fermilab is operated by Fermi Research Alliance, LLC, under Contract No. DE- AC02-07CH11359 with the US Department of Energy. MA acknowledges support by the National Science Foundation under grants OPP-0236449 and PHY-0236449. AT acknowledges a Schroedinger fellowship at DIAS.


\bibliographystyle{h-physrev}
\bibliography{bib}

\end{document}